\documentclass[prd,superscriptaddress,nofootinbib,amsmath,amssymb,aps,11pt]{revtex4-2}

\usepackage{bm}
\usepackage{amsfonts}
\usepackage{latexsym}
\usepackage[latin1]{inputenc}
\usepackage{graphicx}
\usepackage{amsmath}
\usepackage{palatino}
\usepackage{mathpazo}
\usepackage[normalem]{ulem}
\usepackage{xcolor}

\usepackage{booktabs}
\usepackage{dcolumn}

\usepackage{natbib}

\def\jnl@style{\it}
\def\aaref@jnl#1{{\jnl@style#1}}

\def\aaref@jnl#1{{\jnl@style#1}}

\def\aj{\aaref@jnl{AJ}}                   
\def\apj{\aaref@jnl{ApJ}}                 
\def\apjl{\aaref@jnl{ApJ}}                
\def\apjs{\aaref@jnl{ApJS}}               
\def\apss{\aaref@jnl{Ap\&SS}}             
\def\aap{\aaref@jnl{A\&A}}                
\def\aapr{\aaref@jnl{A\&A~Rev.}}          
\def\aaps{\aaref@jnl{A\&AS}}              
\def\mnras{\aaref@jnl{Mon.~Not.~Roy.~Astron.~Soc.}}             
\def\prd{\aaref@jnl{Phys.~Rev.~D}}        
\def\prc{\aaref@jnl{Phys.~Rev.~C}}  
\def\prl{\aaref@jnl{Phys.~Rev.~Lett.}}    
\def\qjras{\aaref@jnl{QJRAS}}             
\def\skytel{\aaref@jnl{S\&T}}             
\def\ssr{\aaref@jnl{Space~Sci.~Rev.}}     
\def\zap{\aaref@jnl{ZAp}}                 
\def\nat{\aaref@jnl{Nature}}              
\def\aplett{\aaref@jnl{Astrophys.~Lett.}} 
\def\apspr{\aaref@jnl{Astrophys.~Space~Phys.~Res.}} 
\def\physrep{\aaref@jnl{Phys.~Rep.}}      
\def\physscr{\aaref@jnl{Phys.~Scr}}       
\def\commat{\aaref@jnl{Comm.~Math.~Phys.}}              
\def\science{\aaref@jnl{Science}}               
\def\cqg{\aaref@jnl{Classical Quant.~Grav.}}            
\def\jpcs{\aaref@jnl{JPCS}}                                     
\def\ijmpd{\aaref@jnl{Int.~J.~Mod.~Phys.~D}}                    
\def\grg{\aaref@jnl{Gen.~Relat.~Gravit.}}               
\def\rpp{\aaref@jnl{Rep.~Prog.~Phys.}}          
\def\npa{\aaref@jnl{Nucl.~Phys.~A}}        
\def\lrr{\aaref@jnl{Living Rev.~Rel.}}                   
\def\jcap{\aaref@jnl{J.~Cosmology Astropart.~Phys.}}    
\def\rmp{\aaref@jnl{Rev.~Mod.~Phys.}}   

\allowdisplaybreaks[1]

\begin{document}

\title{ Neutron stars in extended scalar-Gauss-Bonnet gravity: the richness of the solution spectrum}

\author{Kalin V. Staykov}
\email{kstaykov@phys.uni-sofia.bg}
\affiliation{Department of Theoretical Physics, Faculty of Physics, Sofia University, Sofia 1164, Bulgaria}

\author{Peter Y. Yordanov}
\email{pyordanov@phys.uni-sofia.bg}
\affiliation{Department of Theoretical Physics, Faculty of Physics, Sofia University, Sofia 1164, Bulgaria}

\author{Daniela D. Doneva}
\email{daniela.doneva@uni-tuebingen.de}
\affiliation{Theoretical Astrophysics, Eberhard Karls University of T\"ubingen, T\"ubingen 72076, Germany}
\affiliation{INRNE - Bulgarian Academy of Sciences, 1784  Sofia, Bulgaria}

\author{Stoytcho S. Yazadjiev}
\email{yazad@phys.uni-sofia.bg}
\affiliation{Department of Theoretical Physics, Faculty of Physics, Sofia University, Sofia 1164, Bulgaria}
\affiliation{Institute of Mathematics and Informatics, 	Bulgarian Academy of Sciences, 	Acad. G. Bonchev St. 8, Sofia 1113, Bulgaria}


\begin{abstract}

Neutron stars are natural laboratories for testing gravity in the strong field regime. That is why the full spectrum of neutron star solutions in different modified theories should be thoroughly studied. Among the most natural modifications of general relativity are the theories in which additional scalar degrees of freedom are present. That is why scalar-tensor theories like Brans--Dike and Damour--Esposito--Farese theories, as well as their extensions such as scalar-Gauss-Bonnet gravity, attracted attention throughout the years. In the present work, we combine those theory families and explore extensively the neutron star solution space in their realm. We identify qualitative new behavior of the solutions, including the existence of new types of phase transitions and new branches of solutions present only for high neutron star masses. Due to the peculiarities of the solutions, they can not be easily mimicked by a simple change of the equation of state.
\end{abstract}


\maketitle

\section{Introduction}
Neutron stars (NSs) are among the most compact objects known to exist in the Universe. The spacetime curvature reaches large values for which strong gravity effects become non-negligible. This makes them important ``laboratories'' not only for testing the behavior of matter at extreme densities but also for probing different modifications of general relativity (GR). Despite being often disregarded as a test of strong gravity due to the presence of matter and the associated uncertainties, NS observations are very rich and led to some of the strongest constraints up to date. This includes the binary pulsar constraints \cite{Damour:1991rd,Freire:2012mg,Anderson:2019eay,Yordanov:2024lfk}, the constraints on the speed of gravitational waves through binary mergers \cite{Perkins:2021mhb}, and other multimessenger gravitational tests \cite{Lagos:2024boe,Silva:2020acr}.

That being said, the spectrum of neutron star solutions in different modifications of GR is much less studied compared to the black hole case \cite{Olmo:2019flu,Tuna:2022qqr}. The main efforts, especially in the direction of astrophysical applications, have been put into studying neutron stars in the famous Damour-Esposito-Farese (DEF) model \cite{Damour:1992we,Damour:1993hw}, which is a subclass of the scalar-tensor theories. Other models of neutron stars in modified gravity show, though, that the behavior of solutions can be qualitatively very different \cite{Pani:2011xm,Kleihaus:2016dui,Doneva:2017duq,Charmousis:2021npl,Yagi:2013mbt}. For example, a neutron star with a given baryon mass can be either more or less compact compared to GR, while remaining energetically favorable. Very interesting effects, such as first-order-like phase transitions between different neutron star gravitational phases, can also be observed \cite{Kuan:2022oxs,Doneva:2023kkz}. 

On the other hand, the degeneracies between modified gravity effects and effects coming from the uncertainties in the high-density nuclear matter equation of state (EoS) hinder us from gaining full confidence in our attempt to constrain the behavior of matter through neutron star observations -- one can always speculate that an interesting observational effect is due to a lack of understanding of gravity in the regime of strong spacetime curvature \cite{Shao:2019gjj}. That is why exploring the neutron star solution spectrum in modified gravity in its full complexity, and especially identifying qualitatively new features that can not be easily mistaken with equation of state effects, is an important step. Advancing in this direction is one of the main goals of the present paper.

Our main focus will be the scalar-tensor type of extensions of general relativity (GR) where the gravitational field is mediated not only by the spacetime metric but also by an additional dynamical scalar field. Such theories are not only well motivated by the attempts to quantize gravity, string theory, higher dimensional gravity, etc. Even more importantly, they also offer an important (technically manageable) playgroup for studying possible deviations from GR in the strong field regime that can be used as a guide for designing theory-agnostic tests later on.  A very popular class of the scalar-tensor modifications of  GR are the quadratic theories of gravity where the usual Einstein-Hilbert action is supplemented with all possible algebraic curvature invariants of second order with a dynamical scalar field nonminimally coupled to these invariants \cite{Berti:2015itd}. Such high-curvature corrections are believed to be the low-energy representation of a hypothetical quantum gravity theory. Of particular interest are the quadratic theories where the scalar field is coupled to a special combination of the curvature invariants, namely the Gauss-Bonnet (GB) invariant \cite{Kanti:1995vq,Berti:2015itd}. In this case,  the field equations are of second order as in GR and the theory is free from ghosts. 

The mentioned  properties of scalar-Gauss-Bonnet (sGB) gravity, combined with its quantum gravity motivation and connection to string theory, ignited significant interest during the past years. Contrary to the classical scalar-tensor theories, such as the DEF model, both neutron stars \cite{Pani:2011xm,Saffer:2021gak,Kleihaus:2016dui} and black holes \cite{Kanti:1995vq,Antoniou:2017hxj,Kleihaus:2015aje,Collodel:2019kkx,Cunha:2016wzk,Blazquez-Salcedo:2017txk} can be endowed with a scalar field, exhibiting also the interesting effect of spontaneous scalarization \cite{Doneva:2017bvd,Silva:2017uqg,Antoniou:2017acq,Doneva:2017duq,Cunha:2019dwb} (for a review see \cite{Doneva:2022ewd}). Therefore, in scalar Gauss Bonnet (sGB) gravity, black hole no-scalar-hair theorems can be circumvented. In the classical version of sGB, such as the Einstein-dilaton-Gauss-Bonnet gravity and theories admitting scalarization, the scalar field couples only to the Gauss-Bonnet invariant. This is among the simplest setups that can produce a viable beyond-GR theory with interesting strong field manifestations. 

We shall consider here an extended version of the standard sGB gravity by adding a conformal coupling to matter in the Einstein frame described by a conformal coupling function $A(\varphi)$  in complete analogy with the classical scalar-tensor theories.  We will refer to the model under consideration as extended scalar-tensor theories (ESTT). A similar setup in the context of binary post-merger gravitational wave signal, where the additional (Einstein frame) matter conformal coupling leads to the presence of an additional gravitational wave polarization, namely breathing modes, was considered in \cite{Evstafyeva:2022rve}.
When the GB coupling function is constant, due to the topological nature of the GB invariant, the ESTT under consideration reduces to the classical scalar-tensor theories (STT) with a conformal factor $A(\varphi)$. On the other hand, when $A(\varphi)=1$  we recover the standard sGB gravity. In some sense, the ESTT can be considered as a nonlinear combination of the standard sGB gravity and the classical STT. Neutron star solutions in both limits were extensively studied in the literature. However, the combination of both is poorly examined. With this motivation in mind, in the present paper, we study spherically symmetric neutron stars in various flavors. 

The paper is structured as follows: In Section II we present the basic mathematical constructions and the field equations. In Section III we discuss the choice of coupling functions and present the numerical results in Section IV. The paper ends with Conclusions. 

\section{Mathematical formulation of the theory and regularity condition}
We consider  extended scalar-tensor theories (ESTT) with the following Einstein frame action 
\begin{eqnarray}\label{action}
S=&&\frac{1}{16\pi}\int d^4x \sqrt{-g} 
\Big[R - 2\nabla_\mu \varphi \nabla^\mu \varphi  - V(\varphi)
+ \lambda^2 f(\varphi){\cal R}^2_{GB} \Big] + S_{\rm matter}  ({\cal A}^2(\varphi) g_{\mu\nu},\chi) ,\label{eq:quadratic}
\end{eqnarray}
where $R$ denotes the Ricci scalar with respect to the Einstein frame metric $g_{\mu\nu}$, $\nabla_{\mu}$ is the covariant derivative with respect to  $g_{\mu\nu}$ and  $f(\varphi)$ is the GB coupling function for the scalar field $\varphi$ and $V(\varphi)$ is its potential. The GB coupling constant $\lambda$ has dimension of $length$ and ${\cal R}^2_{GB}$ denotes the Gauss-Bonnet invariant defined by ${\cal R}^2_{GB}=R^2 - 4 R_{\mu\nu} R^{\mu\nu} + R_{\mu\nu\alpha\beta}R^{\mu\nu\alpha\beta}$ where $R_{\mu\nu}$ and $R_{\mu\nu\alpha\beta}$ are the Ricci tensor and  the Riemann tensor respectively with respect to the Einstein frame metric $g_{\mu\nu}$. $S_{\rm matter}$ is the Einstein frame matter action where the matter fields are collectively denoted by $\chi$ and $A(\varphi)$ is the Einstein frame conformal  coupling function between the scalar field and the matter. A similar type of GB ESTT has already been considered in the direction of binary mergers \cite{Evstafyeva:2022rve} because, contrary to the classical sGB gravity without (Einstein frame) conformal matter coupling, it can give rise to the presence of additional polarization of the gravitational field, namely breathing modes induced by the scalar radiation.

The physical metric is the Jordan frame metric $\tilde{g}_{\mu\nu}$ is given by $\tilde{g}_{\mu\nu}=A^2(\varphi) g_{\mu\nu}$. Furthermore, the Einstein frame energy-momentum tensor $T_{\mu\nu}$ is related to the Jordan frame one $\tilde{T}_{\mu\nu}$ through the relation ${\tilde T}_{\mu\nu}=A^2(\varphi) T_{\mu\nu}$. We should note that the considered theory, when transformed to the physical Jordan frame, shares some similarities with the Ricci coupling Gauss-Bonnet gravity examined in \cite{Antoniou:2021zoy,Ventagli:2021ubn} because the matter conformal coupling will transform into a Ricci scalar field coupling. They are not equivalent, though, since the GB invariant will transform as well in a nontrivial way. 

In the particular case when $f(\varphi)=const$ we recover the classical scalar-tensor theories while for $A(\varphi)=1$ we have the standard scalar-Gauss-Bonnet (sGB) gravity. Thus, the extended
scalar-tensor theories given by the action in eq. (\ref{action}) can be considered as a sort of nonlinear combination between the classical scalar-tensor theories and the standard 
sGB gravity. In what follows we consider the case with $V(\varphi)=0$ for simplicity.

In the present paper, we are interested in static and spherically symmetric neutron star solutions in asymptotically flat spacetime. Hence we consider general static and spherically symmetric ansatz for the Einstein frame metric
\begin{eqnarray}
ds^2= - e^{2\Phi(r)}dt^2 + e^{2\Lambda(r)} dr^2 + r^2 (d\theta^2 + \sin^2\theta d\phi^2 ).
\end{eqnarray}   

The matter source is assumed to be a perfect fluid with $T_{\mu\nu}=(\rho + p)u_{\mu}u_{\nu} + pg_{\mu\nu}$ 
where $\rho$, $p$ and $u^{\mu}$ are the Einstein frame energy density, pressure, and 4-velocity of the fluid, respectively. We also require the perfect  
fluid and the scalar field to respect the staticity and the spherical symmetry. With these
conditions imposed, the dimensionally reduced field equations are:
\begin{eqnarray}
&&\frac{2}{r}\left[1 +  \frac{2}{r} (1-3e^{-2\Lambda})  \Psi_{r}  \right]  \frac{d\Lambda}{dr} + \frac{(e^{2\Lambda}-1)}{r^2} 
- \frac{4}{r^2}(1-e^{-2\Lambda}) \frac{d\Psi_{r}}{dr} - \left( \frac{d\varphi}{dr}\right)^2=8\pi {\cal A}^4(\varphi)\tilde{\rho} e^{2\Lambda}, \label{DRFE1}\\ && \nonumber \\
&&\frac{2}{r}\left[1 +  \frac{2}{r} (1-3e^{-2\Lambda})  \Psi_{r}  \right]  \frac{d\Phi}{dr} - \frac{(e^{2\Lambda}-1)}{r^2} - \left( \frac{d\varphi}{dr}\right)^2=8\pi {\cal A}^4(\varphi)\tilde{p} e^{2\Lambda},\label{DRFE2}\\ && \nonumber \\
&& \frac{d^2\Phi}{dr^2} + \left(\frac{d\Phi}{dr} + \frac{1}{r}\right)\left(\frac{d\Phi}{dr} - \frac{d\Lambda}{dr}\right)  + \frac{4e^{-2\Lambda}}{r}\left[3\frac{d\Phi}{dr}\frac{d\Lambda}{dr} - \frac{d^2\Phi}{dr^2} - \left(\frac{d\Phi}{dr}\right)^2 \right]\Psi_{r} 
\nonumber \\ 
&& \hspace{0.5cm} - \frac{4e^{-2\Lambda}}{r}\frac{d\Phi}{dr} \frac{d\Psi_r}{dr} + \left(\frac{d\varphi}{dr}\right)^2=8\pi {\cal A}^4(\varphi)\tilde{p} e^{2\Lambda}, \label{DRFE3}\\ && \nonumber \\
&& \frac{d^2\varphi}{dr^2}  + \left(\frac{d\Phi}{dr} \nonumber - \frac{d\Lambda}{dr} + \frac{2}{r}\right)\frac{d\varphi}{dr} \nonumber \\ 
&& \hspace{0.5cm} - \frac{2\lambda^2}{r^2} \frac{df(\varphi)}{d\phi}\Big\{(1-e^{-2\Lambda})\left[\frac{d^2\Phi}{dr^2} + \frac{d\Phi}{dr} \left(\frac{d\Phi}{dr} - \frac{d\Lambda}{dr}\right)\right]    + 2e^{-2\Lambda}\frac{d\Phi}{dr} \frac{d\Lambda}{dr}\Big\} = \nonumber\\ && \hspace{0.5cm} 4\pi \alpha(\varphi){\cal A}^4(\varphi)(\tilde{\rho} - 3\tilde{p})e^{2\Lambda}, \label{DRFE4}
\end{eqnarray}
where  

\begin{eqnarray}
\Psi_{r}=\lambda^2 \frac{df(\varphi)}{d\varphi} \frac{d\varphi}{dr}.
\end{eqnarray}

The equation for hydrostatic equilibrium of the fluid has the form
\begin{eqnarray}\label{FEE}
\frac{d\tilde{p}}{dr} = - (\tilde{\rho} + \tilde{p}) \left(\frac{d\Phi}{dr} + \alpha(\varphi) \frac{d\varphi}{dr}\right),
\end{eqnarray}  
where the function $\alpha(\varphi)$ is defined as 
\begin{equation}
	\alpha(\varphi) = \frac{d \ln{{\cal A}(\varphi)}}{d\varphi}.
\end{equation}
In the above equations the tilde quantities are with respect to the (physical) Jordan frame, and they transform between the two frames in the following way 
\begin{eqnarray}\label{DPTEJF}
	\rho &=&{\cal A}^4(\varphi){\tilde\rho}, \nonumber \\
	p&=&{\cal A}^4(\varphi){\tilde p},  \\
	u_{\mu}&=& {\cal A}^{-1}(\varphi){\tilde u}_{\mu}. \nonumber
\end{eqnarray}
The Jordan frame radius of the star is defined as $R_s = {\cal A}(\varphi_s)r_s$, where $r_s$ is the radial coordinate in Einstein frame at which the pressure vanishes $\tilde{p}(r_s) = 0$.

Using the expansion of the field equations close to the center of the star one can derive important relations determining the existence of solutions. If one substitutes\footnote{Note that even though $A(\varphi)$ and $\alpha(\varphi)$ are functions of the scalar field $\varphi$, we expand them independently in order to keep them in most general form, without assuming any analytical expression.}
\begin{eqnarray}
&&\Lambda=\Lambda_0 + \Lambda_1 r+\frac{1}{2} \Lambda_2 r^2  + O(r^3), \;\;\; \Phi=\Phi_0+\Phi_1 r+\frac{1}{2} \Phi_2 r^2 + O(r^3), \nonumber \\ &&  \varphi=\varphi_0+\varphi_1 r + \frac{1}{2} \varphi_2 r^2 + O(r^3), \;\;\; \tilde{p} = \tilde{p_0} + \tilde{p_1} r +  \frac{1}{2} \tilde{p_2} r^2 + O(r^3), \nonumber \\ && \tilde{\rho} = \tilde{\rho_0} + \tilde{\rho_1} r +  \frac{1}{2} \tilde{\rho_2} r^2 + O(r^3), \;\;\; \alpha(\varphi) = \alpha_0 + \alpha_1 r + \frac{1}{2} \alpha_2 r^2 + O(r^3), \nonumber \\ && A(\varphi) = A_0 + A_1\varphi_1 r + \frac{1}{2} (A_2\varphi_1^2 + A_1 \varphi_2) r^2 + O(r^3)
\end{eqnarray}
and expands the differential equations \eqref{DRFE1}--\eqref{DRFE4} around the center of the star then it is easy to obtain that $\Lambda_0=0$, $\Lambda_1=0$, $\Phi_1=0$ and $\varphi_1=0$. In addition, the following condition should also be  fulfilled
\begin{eqnarray}
&&9\,{\Lambda_2}^{4}{\left(\frac{df(\varphi_0)}{d\varphi}\right)}^{2}{\lambda}^{4}+72\,{{A_0}}^{4}{
\Lambda_2}^{3}\pi\,{\left(\frac{df(\varphi_0)}{d\varphi}\right)}^{2}{\lambda}^{4}\tilde{p}_0+\\ \nonumber && 32\,
\, \left( \pi\,\alpha_0\,{\lambda}^{2}\left(\frac{df(\varphi_0)}{d\varphi}\right)\, \left( \tilde{\rho}_0 - 3\,\tilde{p}_0 \right) {{A_0}}^{4}   -3/16\right) \pi\tilde{\rho}_0\,{{ A_0}}^{4}
\Lambda_2+16\,{{A_0}}^{8}{\pi}^{2}{\tilde{\rho}_0}^{2}
 = 0
\end{eqnarray}

This equation serves the purpose of an existence condition, namely real roots of this quartic algebraic equation (with respect to $\Lambda_2$) should exist in order for the regularity at the neutron star center to be satisfied. A violation of this condition typically happens for large neutron star masses and strong scalar fields. Thus, the solution branches are terminated at a certain central energy density, as we will see below.

\section{Scalar field coupling choices}
We will present exhaustive combinations of couplings in the classical scalar-tensor theory and sGB gravity. Some of them will be simple deformations while others will lead to significant qualitative change in the solution spectrum. 
Regarding the sGB theories, we consider:
\begin{itemize}
    \item \textbf{sGB-shift-sym:} Shift-symmetric sGB gravity with a linear scalar field  coupling. This theory is characterized by the fact that it always deviates from GR since $\varphi=0$ is not a solution of the field equations\footnote{In this paper we consider the case when scalar field is zero at infinity}. Neutron stars always have zero scalar charge \cite{Yagi:2015oca}.
    \begin{equation}
        f_{\rm Sym}(\varphi) = \varphi.
    \end{equation}
    \item \textbf{sGB-EdGB:} Einstein-dilaton-Gauss-Bonnet (EdGB) gravity. In that case, $\varphi=0$ is not a solution of the field equations. As a matter of fact, the shift-symmetric case can be considered as a weak coupling limit of the EdGB gravity, since the former represents the first term in the EdGB coupling function expansion for weak scalar fields. An interesting qualitative difference between the two, though, is that in EdGB gravity neutron stars have a nonzero scalar charge.
    \begin{equation}
        f_{\rm EdGB}(\varphi) = \frac{1}{2\beta_{\rm EdGB}}e^{2\beta_{\rm EdGB} \varphi}.
    \end{equation}
    \item \textbf{sGB-SS:} sGB gravity with a  coupling function admitting spontaneous scalarization. Hence, it fulfills the conditions $\frac{df}{d\varphi}(0)=0$ and $\frac{d^2f}{d\varphi^2}(0)\ne 0$. 
    \begin{equation}
        f_{\rm SS}(\varphi)=  \pm \frac{1}{2\beta_{\rm SS}} \left[1-e^{-\beta_{\rm SS}\varphi^2}\right]. \label{eq:sGB_scalarize}
    \end{equation}
    The two signs in front of the coupling lead to an important difference -- both neutron stars and black holes can scalarize (the ``$+$'' sign), or black holes can not carry scalar hair and only neutron stars can scalarize (the ``$-$'' sign).
\end{itemize}

As far as the scalar-tensor theories are concerned, we consider the two most standard cases -- the Brans-Dicke and the  Damour-Esposito-Farese models
\begin{itemize}
	\item \textbf{STT-BD}: Brans-Dicke theory with the following conformal coupling: 
	\begin{equation}\label{eq:Amatter2}
		{\cal A}_{\rm BD}(\varphi) = e^{\gamma_{\rm BD}\varphi}, 	\alpha_{\rm BD}(\varphi) = \gamma_{\rm BD}.
	\end{equation}
 In that case, $\varphi=0$ is not a solution of the field equations and the theory is severely constrained from weak field observations and binary pulsars \cite{Will_2018,Freire:2024adf}. Nevertheless, this theory is of particular interest, because it allows for the presence of the so-called breathing modes
\item \textbf{STT-DEF}: Damour-Esposito-Farese model where $\varphi=0$ is a solution of the field equations and spontaneous scalarization of neutron stars is observed \cite{Damour:1992we} 
\begin{equation} \label{eq:Amatter1}
	{\cal A}_{\rm DEF}(\varphi) = e^{\frac{1}{2}\gamma_{\rm DEF}\varphi^2}, \alpha_{\rm DEF}(\varphi) = \gamma_{\rm DEF} \varphi.
\end{equation}
\end{itemize}
Clearly, a number of different combinations can be formed from the above mentioned coupling functions. In the main text of the paper we will present only the most interesting ones while the rest will be given in Appendices.

Let us comment on the observational constraints one can impose on the parameters in the coupling functions above. The weak field observations \cite{Will_2018} as well as the up to date binary pulsar observations have already set tight limits on the coupling parameters. Namely, the Brans-Dicke parameter is constrained down to roughly $|\gamma_{\rm BD}| < 10^{-3}$ \cite{Will_2018,Freire:2024adf}. For the shift-symmetric Gauss-Bonnet theory the latest constraints are $\lambda_{\rm Sym} < 3.01$ \cite{Lyu:2022gdr} and for the EdGB with fixed $\beta_{\rm EdGB} = 0.14$   the authors of \cite{Yagi:2015oca} constrain the upper boundary of   $\lambda_{\rm EdGB}$ in the interval between $2.5$ and $5.1$ (all numbers are given in our dimensionless units). A two-dimensional constraint of the   ($\lambda_{\rm EdGB},\beta_{\rm EdGB}$) parameter space was provided in \cite{Yordanov:2024lfk}.

Constraining spontaneously scalarized models is more subtle because this is a strong field effect in its essence. Currently, the best constraints come from the binary pulsar observations (see \cite{Freire:2024adf} and references therein) while the merger of black holes is also promising especially with the next generation of gravitational wave detectors  \cite{Shao:2017gwu}. The observations practically rule out the DEF model \cite{Zhao:2022vig} as well as neutron star scalarization in sGB gravity \cite{Danchev:2021tew,Wong:2022wni}. While the constraints on combinations of sGB gravity and scalar-tensor theories have not been studied yet, one can expect that they are of similar order. 

There is a caveat, though, that justifies considering larger values of the above-mentioned parameters. This is connected to the fact that all of the constraints discussed above are in the massless scalar field case. If one adds  scalar field mass $m_{\varphi}$, then the scalar field acquires an effective radius (its Compton wavelength) above which it drops exponentially. Thus, a number of scalar field manifestations can be hidden inside this effective radius. On the other hand, from a field theory perspective, it is completely justified to expect that a potential new fundamental scalar field would be massive. Little studies of the observational constraints have been performed in the massive scalar field case \cite{Kuan:2023hrh,Tuna:2022qqr} but overall it is expected that at least moderate deviations from GR are admitted. 

The overall influence of a scalar field mass  on the solution spectrum is typically to simply suppress the deviations from GR at distances larger than  the scalar field Compton wavelength,  without altering the neutron star solution structure qualitatively. Therefore, the $V(\varphi)=0$ neutron star models presented below can be viewed as the maximum deviation from GR one can have in the $m_\varphi \rightarrow 0$ limit.

\section{Numerical results}

In this study we are interested in the existence of neutron star solutions and how the combination between the GB coupling $f(\varphi)$  and the Einstein frame conformal matter coupling $A(\varphi)$   will effect their structure and properties. Being only a qualitative study, we concentrate on a single realistic equation of state, namely MPA1 \cite{Muther:1987xaa}, and we employ its piecewise polytropic approximation \cite{Read:2008iy}.  

The system of equations (\ref{DRFE1})-(\ref{DRFE4}), combined with the equation for hydrostatic equilibrium (\ref{FEE}), and supplied with an EoS and proper boundary conditions, is solved numerically. The central values of the metric function $\Phi$ and the scalar field $\varphi$ are obtained via a shooting method.  The boundary conditions are the natural ones, namely regularity at the center of the star:
\begin{equation}
	\left.\Lambda\right|_{r\rightarrow 0}\rightarrow 0,\quad \left. \frac{d\Phi}{dr}\right|_{r\rightarrow 0}\rightarrow 0, \quad \left.\frac{d\varphi}{dr}\right|_{r\rightarrow 0}\rightarrow 0
\end{equation}
and asymptotic flatness at infinity:
\begin{equation}
	\left.\Lambda\right|_{r\rightarrow \infty}\rightarrow 0,\quad \left.\Phi\right|_{r\rightarrow \infty}\rightarrow 0, \quad \left.\varphi\right|_{r\rightarrow \infty}\rightarrow 0.
\end{equation}

An important characteristic of the neutron star solutions is the scalar charge $D$ which is defined through the leading order expansion of the scalar field at   $r\to \infty$
\begin{equation}
	\varphi \simeq \frac{D}{r} + O(1/r^2).
\end{equation}

In what follows, the star parameters will be presented in the physical Jordan frame and we use the dimensionless parameter
\begin{eqnarray}
\lambda \to \frac{\lambda}{R_0}
\end{eqnarray}
where $R_0 = 1.47664$ km which corresponds to one solar mass.

\subsection{Limiting cases of pure scalar-Gauss-Bonnet and pure scalar-tensor theories}
\begin{figure}[htb]
	\includegraphics[width=0.33\textwidth]{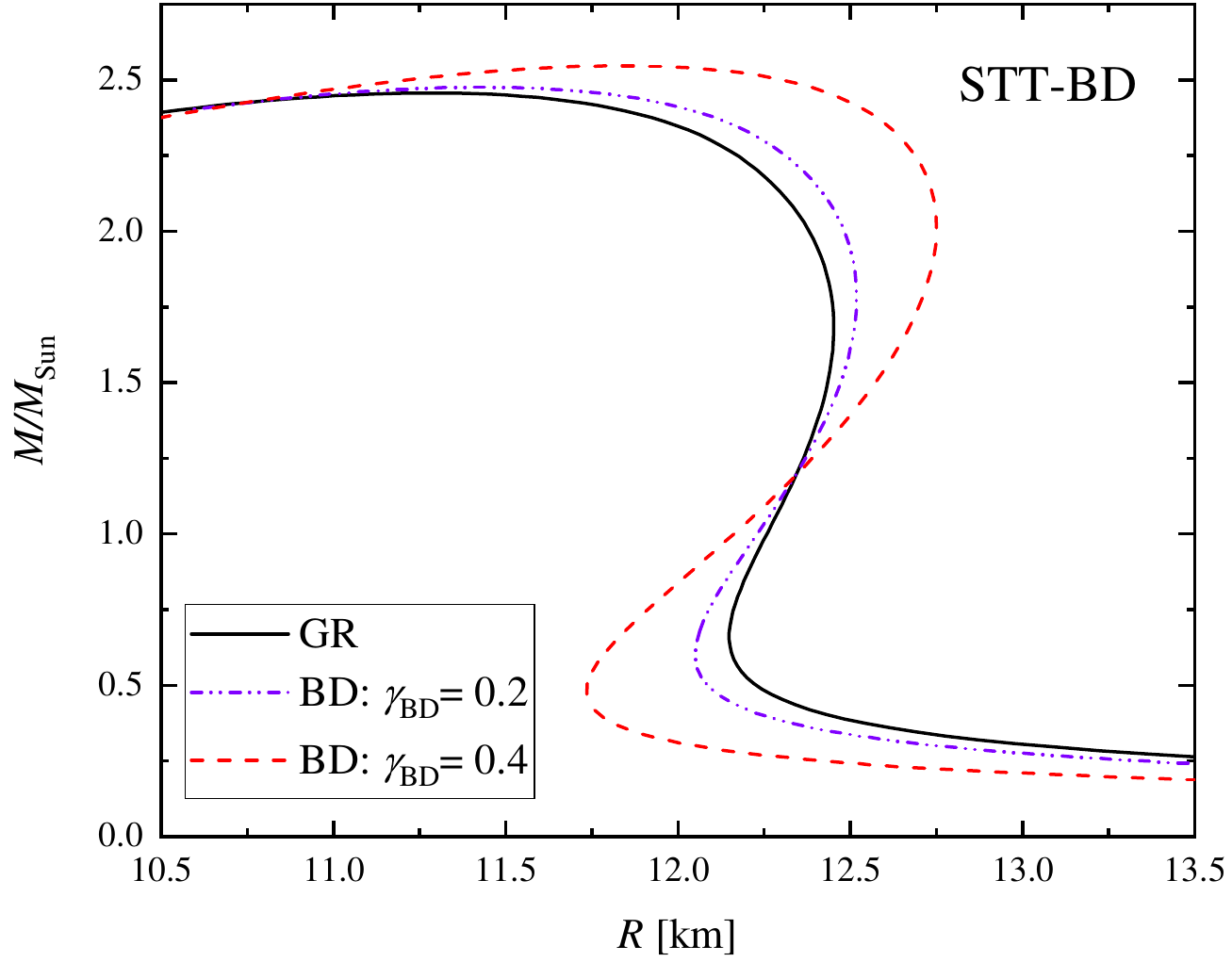}
	\includegraphics[width=0.33\textwidth]{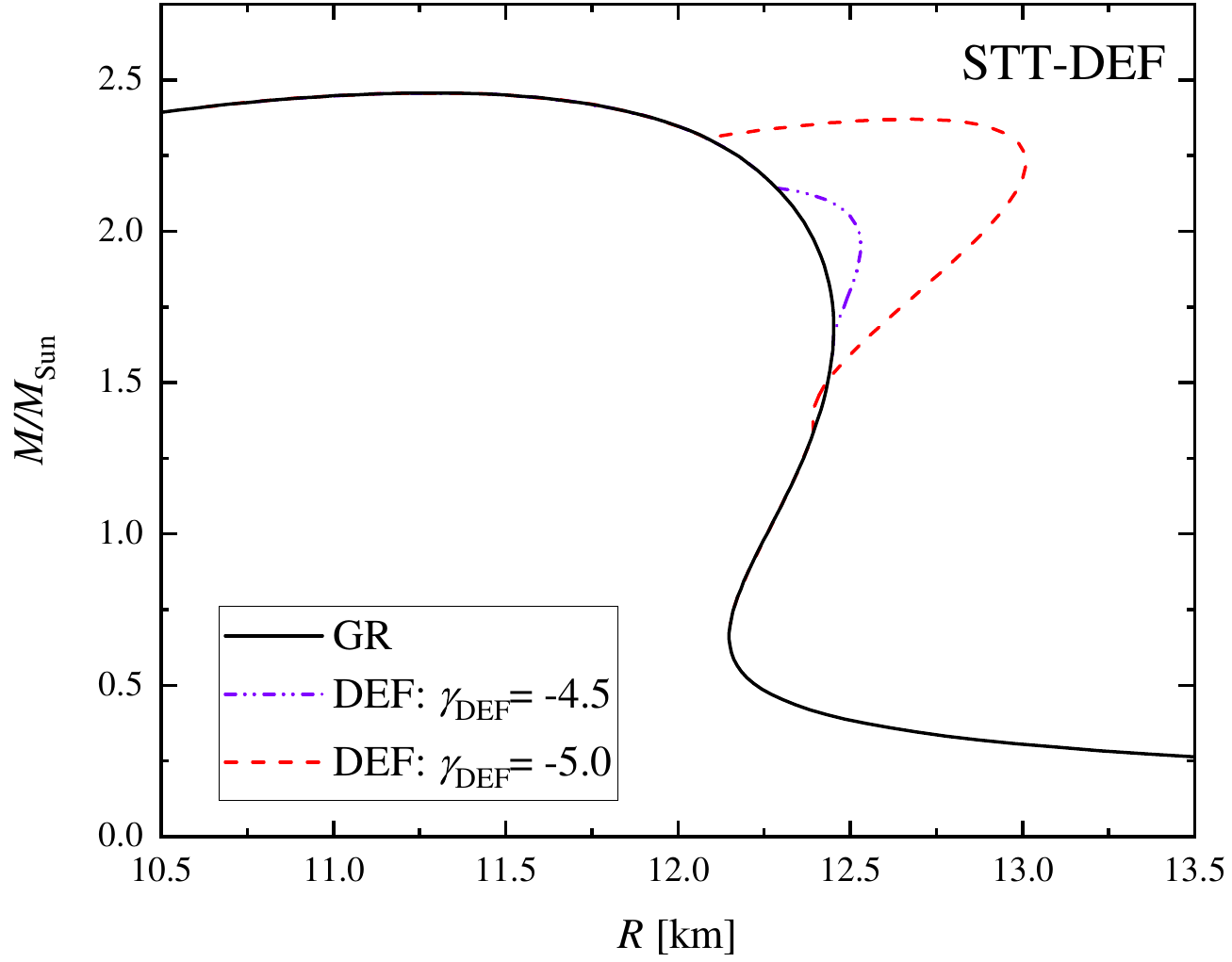}
 	\includegraphics[width=0.33\textwidth]{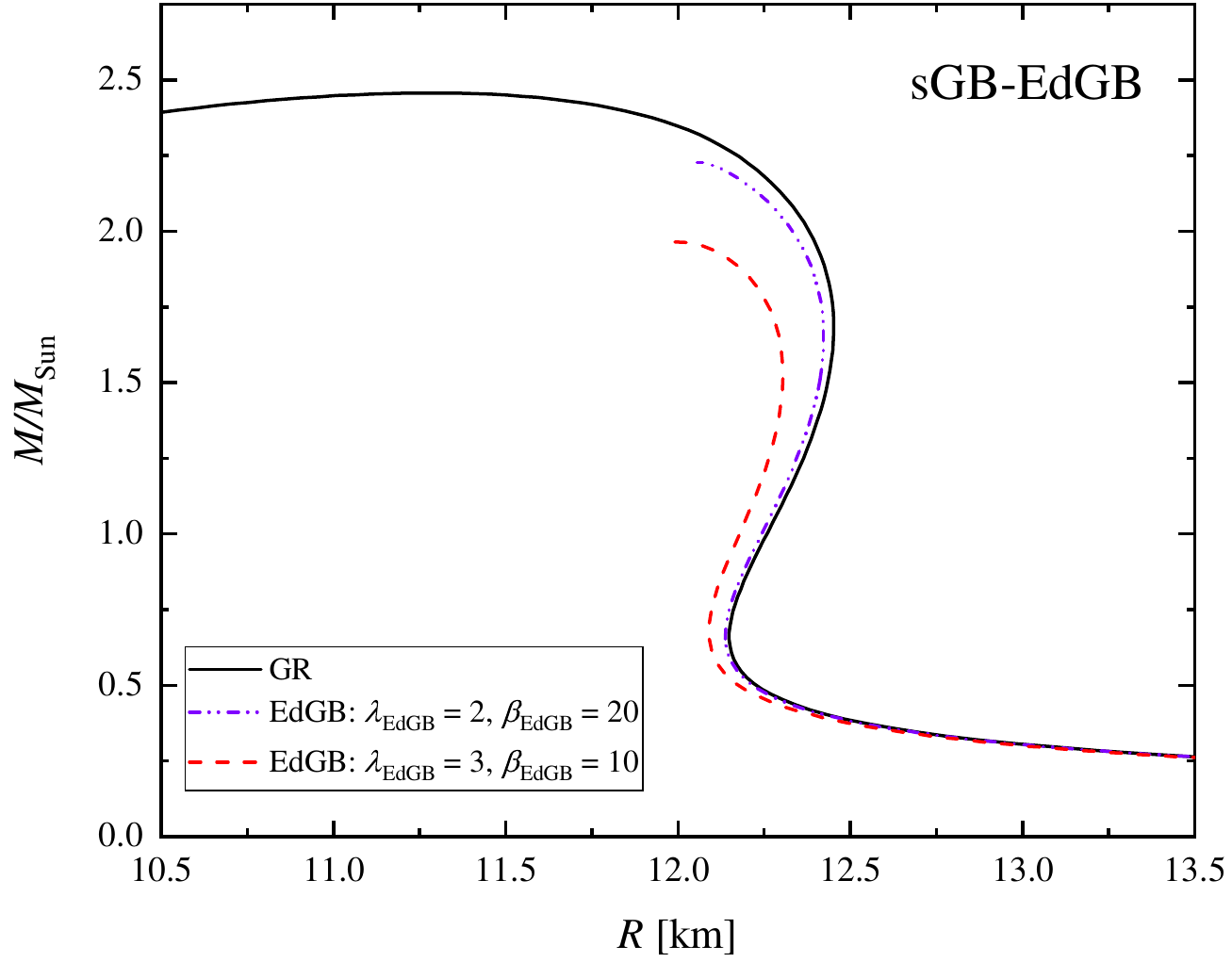}
	\includegraphics[width=0.33\textwidth]{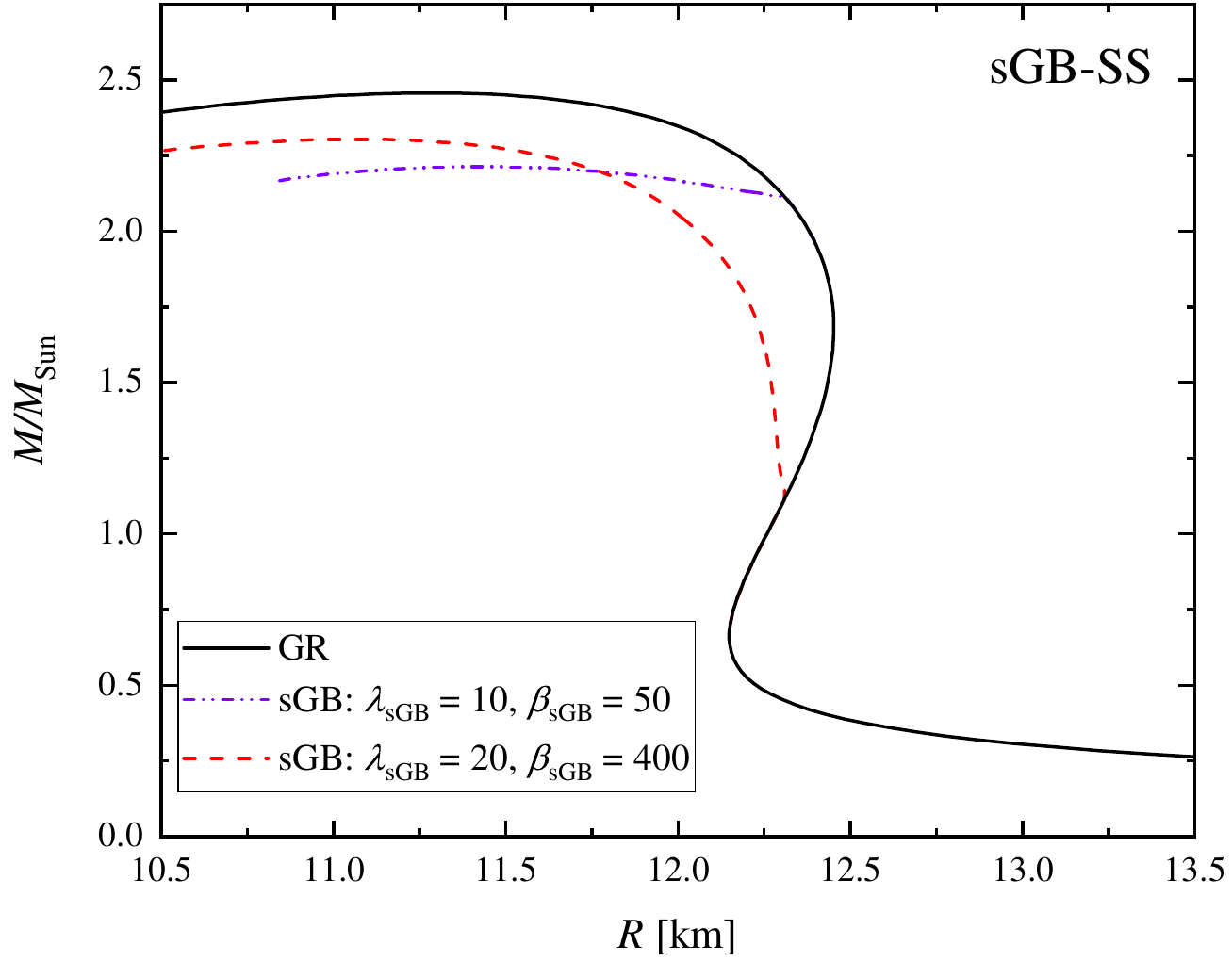}
	\caption{Mass of radius relations for pure STT and pure GB theories. \textit{Top-left} Brans-Dicke scalar-tensor theory; \textit{Top-Right} Damour-Esposito-Farese model; \textit{Bottom-Left} EdGB gravity with exponential coupling; \textit{Top-Right} sGB theory with scalarization.}
	\label{fig:M_R_pure}
\end{figure}

We start first by discussing briefly the limiting cases of pure sGB theory (with or without scalarization), pure Brans-Dicke theory, and DEF model presented in Fig. \ref{fig:M_R_pure}. The idea is to remind the reader the main qualitative differences between them in order to understand better the peculiar behavior of the solution branches exhibited in the next subsections.

In sGB gravity, the neutron stars tend to be more compact compared to GR and their maximum masses are always lower (for a fixed central energy density). Thus the sGB branch is always ``below'' the GR one as seen in the bottom left and right panels of Fig. \ref{fig:M_R_pure}. This applies to all cases, namely the shift-symmetric GB theory, EdGB, and the case admitting scalarization. In scalar-tensor theories, it is the other way around -- the Brans-Dicke model and for the majority of the DEF neutron stars (excluding the low-density part of the scalarized branch), the compactness is smaller while the maximum mass is larger compared to GR. Thus, the scalar-tensor neutron stars are ``above'' the GR ones as seen in the top left and right panels of Fig. \ref{fig:M_R_pure}. The theories where $\varphi=0$ is not a solution of the field equations are characterized by a single neutron star branch (shift symmetric GB, EdGB, and BD STT). In the rest of the cases admitting spontaneous scalarization (sGB gravity with coupling \eqref{eq:sGB_scalarize} and DEF model) we have two branches -- a GR one with $\varphi=0$ and a scalarized one. 

These important qualitative differences between the different couplings considered in the present paper give rise to the very interesting solution structures in the mixed GB-STT case, demonstrated in the next subsections.

\subsection{``sGB-SS $+$ STT-BD'' sGB gravity with spontaneous scalarization combined with Brans-Dicke scalar-tensor theory}
\begin{figure}[htb]
	\includegraphics[width=0.33\textwidth]{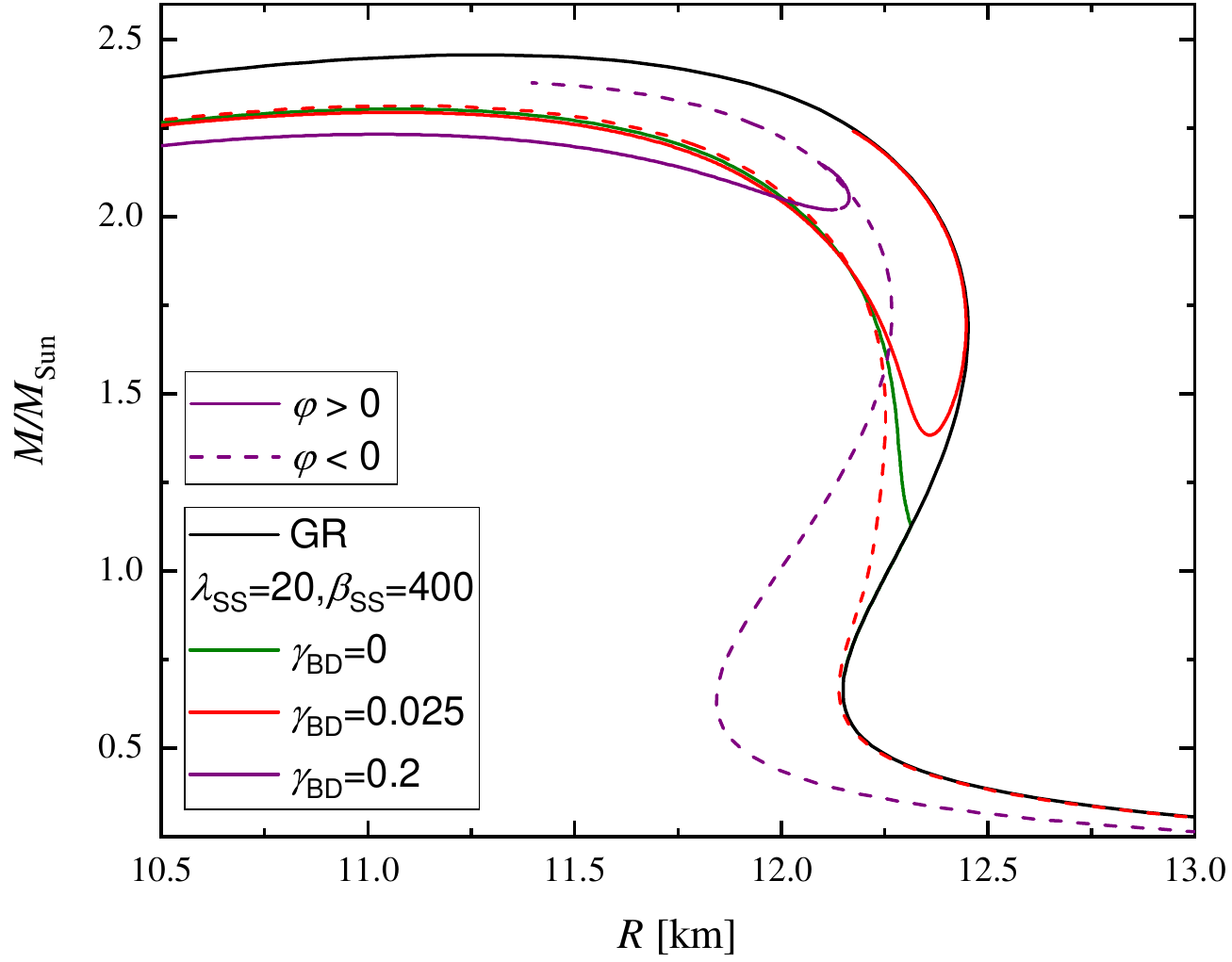}
	\includegraphics[width=0.33\textwidth]{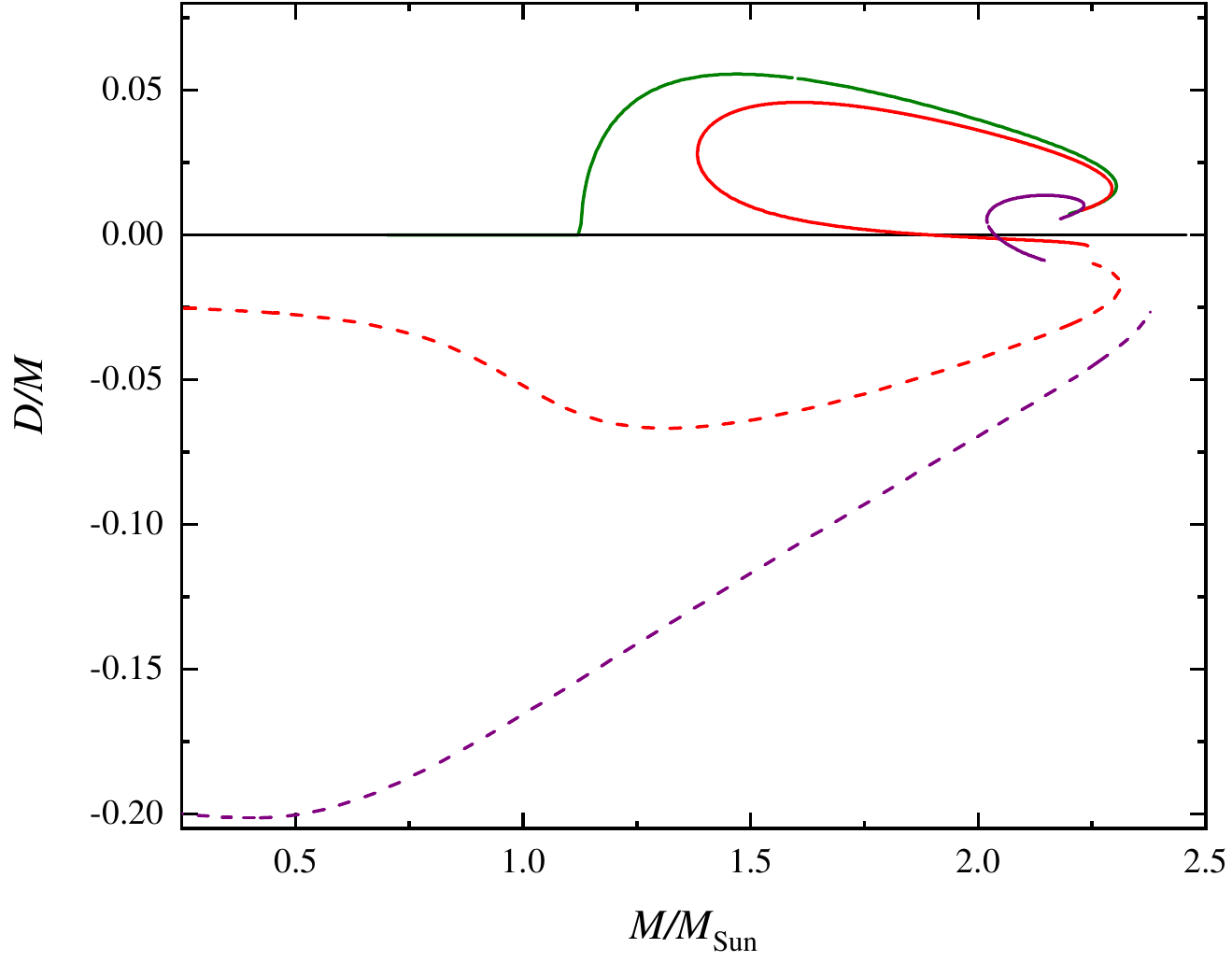}
	\includegraphics[width=0.33\textwidth]{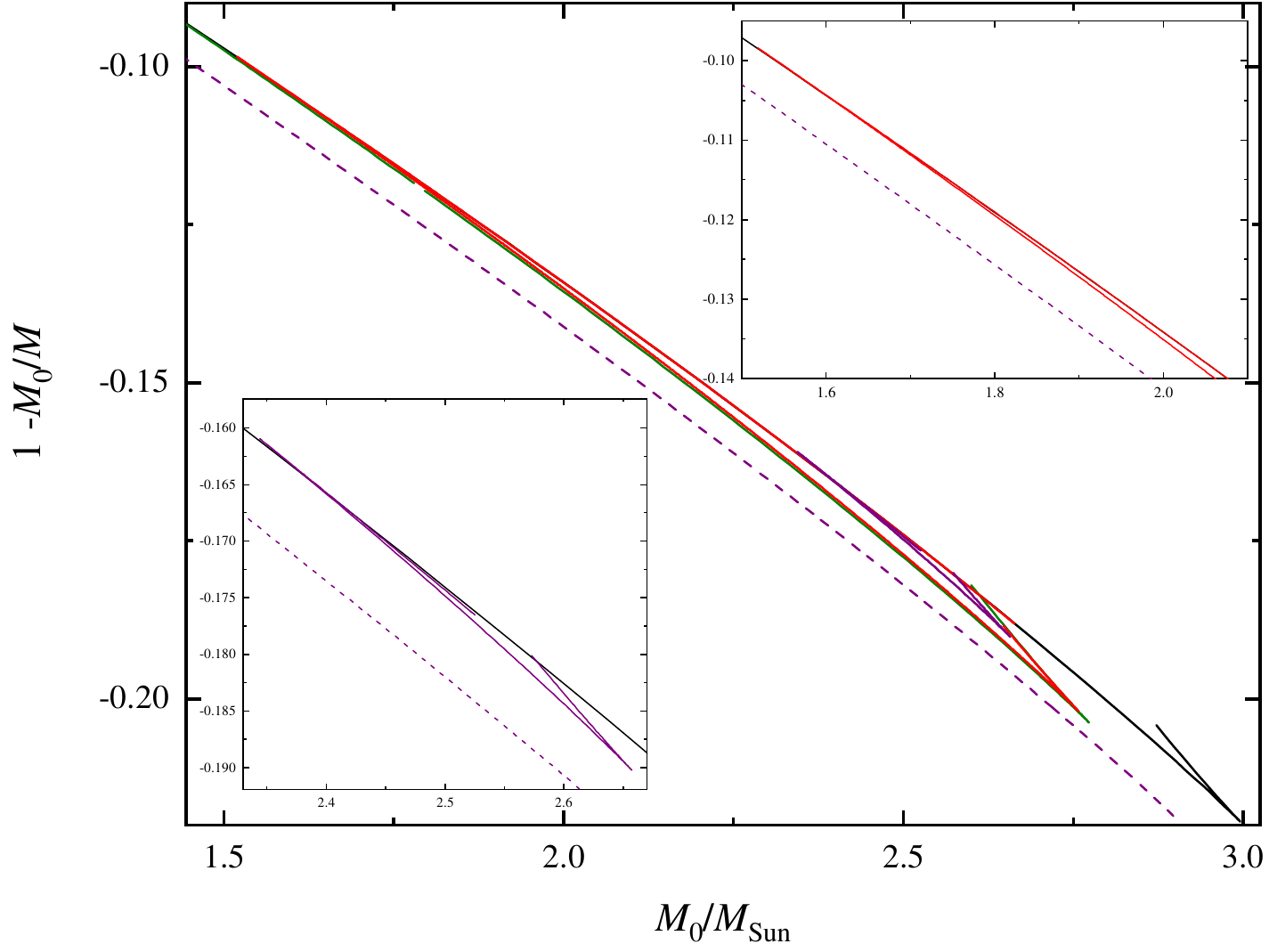}
	\caption{CASE ``sGB-SS $+$ STT-BD'': sGB gravity with spontaneous scalarization combined with BD STT. \textit{Left:} Mass of the star as a function of its radius. \textit{Middle:} The scalar charge of the star, normalized to the stellar mass, as a function of mass. \textit{Right:} The binding energy as a function of the baryon mass. }
	\label{fig:c1_A2}
\end{figure}

The presentation of the results we start with the case we find to be the most interesting, namely the theory defined by the combination of the  following  coupling functions
\begin{equation}
f_{\rm SS}(\varphi)=  \pm \frac{1}{2\beta_{\rm SS}} \left[1-e^{-\beta_{\rm SS}\varphi^2}\right] \label{eq:sGB_coupling}, \,\,\,\,\,\, A_{\rm BD}(\varphi)=e^{\gamma_{\rm BD} \varphi}, 
\end{equation} 
where $\beta_{\rm SS}>0$ is a parameter. Investigations have been carried out for both signs of $f_{\rm SS}(\varphi)$ and the structure of branches is qualitatively very similar. That is why in Fig. \ref{fig:c1_A2} we present results only for the ``$-$'' sign (in this case scalarization of black holes is not possible and only neutron stars can carry scalar field). A similar setup of a sGB gravity with a BD theory, in the form of a conformal coupling of the scalar field to the metric in the matter-field action, was considered in \cite{Evstafyeva:2022rve}. Such a coupling leads to the existence of an additional polarization of the gravitational waves (breathing modes) that is potentially detectable during binary merger events.

For the above combination of coupling functions the theory is not symmetric with respect to the sign on the scalar field. We have a symmetry, though,  if one simultaneously changes the sign of the scalar field and the parameter $\gamma_{\rm BD}$. Depending on whether the scalar field and $\gamma_{\rm BD}$ have identical or opposite signs, we were able to find different types of solutions. Both cases are presented in Fig. \ref{fig:c1_A2} in which we plot the mass of radius relation (left panel), the normalized scalar charge as a function of the stellar mass (middle panel), and the binding energy as a function of the neutron star rest mass (right panel).

Let us discuss first the case with $\varphi$ and $\gamma_{\rm BD}$ being with  opposite signs, i.e. $\varphi \gamma_{\rm BD} < 0$. For very small values of $\gamma_{\rm BD}$ the solution branch closely follows the GR one for small masses (purple dashed line in Fig. \ref{fig:c1_A2}). Around the pure sGB bifurcation point it smoothly deviates from GR and starts to closely follow the pure sGB branch. With the increase of $\gamma_{\rm BD}$ the branch first deviates from both GR and sGB around the pure sGB bifurcation point. With the increase of $\gamma_{\rm BD}$ the low mass models move to higher compactness with respect to GR while the high mass models move to lower compactness with respect to pure sGB and the maximal mass increases.

A second type of solution branches is observed when the scalar field and $\gamma_{\rm BD}$ have  the same sign, i.e. $\varphi \gamma_{\rm BD} > 0$. For small absolutes value of $\gamma_{\rm BD}$ the new solution branch follows the pure sGB one, tending to the bifurcation point (solid red line in Fig. \ref{fig:c1_A2}). At some minimal mass, higher than the bifurcation one, this branch smoothly transitions into a secondary branch and the mass starts to increase while the branch closely follows the GR one. With the increase of the absolute value of $\gamma_{\rm BD}$ (dashed purple line) both branches deviate from sGB and GR correspondingly, to lower masses. The minimal mass at which the transition occurs increases and the secondary branch gets shorter. Naturally, the scalar charge also behaves non-monotonically for part of the branches as seen in the middle panel of Fig. \ref{fig:c1_A2}

The binding energy, presented in the right panel of Fig. \ref{fig:c1_A2} can be used as an indication of the stability. For all scalarized models the binding energy is higher in absolute value, compared to the GR one. The cusps in the figure signal a change of stability. Thus, the branch having larger binding energy (by absolute value) in the vicinity of the cusp, is most probably stable. A cusp is always observed at the maximum mass point for all branches. A second cusp can be observed for the branches having a minimal mass. Based on the arguments above, one can conclude that most probably, the middle part of the branch (between the minimum and maximum mass) is stable while the rest of the branches are unstable. We should note that this is a very unusual situation  and thus a rigorous linear stability analysis should be performed. 

Another question that arises is which one of the potentially stable branches realizes in practice. The most probable answer is the dashed branches having $\varphi \gamma_{\rm BD} < 0$ since they have the largest binding energy (by absolute value). It is not clear yet, though, because at least sp,e of the branches with  $\varphi \gamma_{\rm BD} > 0$ are most probably linearly stable. Thus only a proper dynamical nonlinear evolution of a specific process can give an answer about the end state of a newly formed neutron star via core-collapse or binary merger.

It is important to note, especially related to the minimal mass in the  $\varphi \gamma_{\rm BD} > 0$ case, that the observed behavior of the solutions quantitatively depends on $\lambda_{\rm SS}$ and $\beta_{\rm SS}$. $\lambda_{\rm SS}$ controls the position of the bifurcation point while $\beta_{\rm SS}$ is related to the magnitude of the deviation from GR branch. Therefore, the minimum mass of the  $\varphi \gamma_{\rm BD} > 0$ branches is controlled by $\lambda_{\rm SS}$.

\subsection{``sGB-SS $+$ STT-DEF'': sGB with spontaneous scalarization combined with DEF spontaneous scalarization model.} 

\begin{figure}[htb]
	\includegraphics[width=0.33\textwidth]{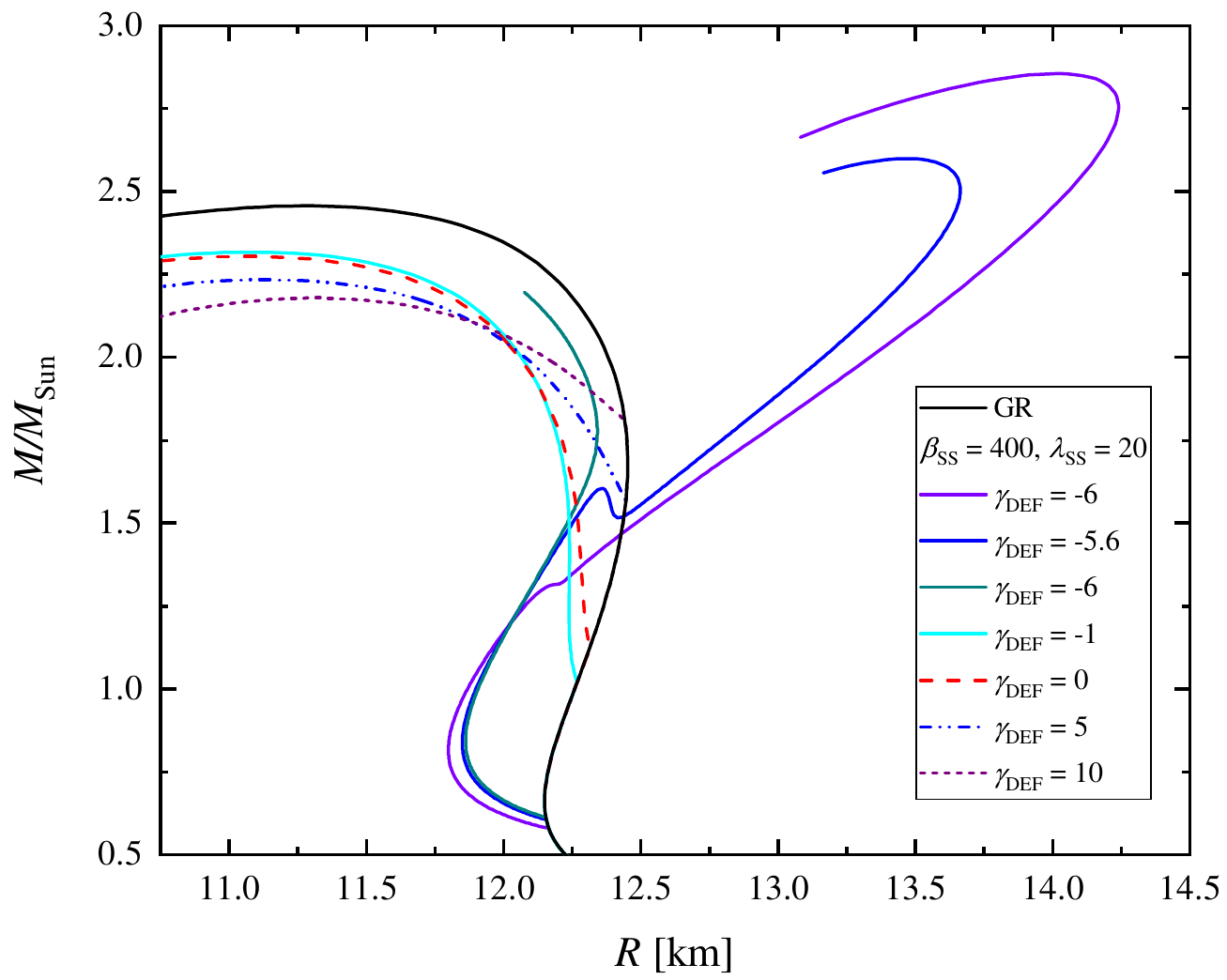}
	\includegraphics[width=0.33\textwidth]{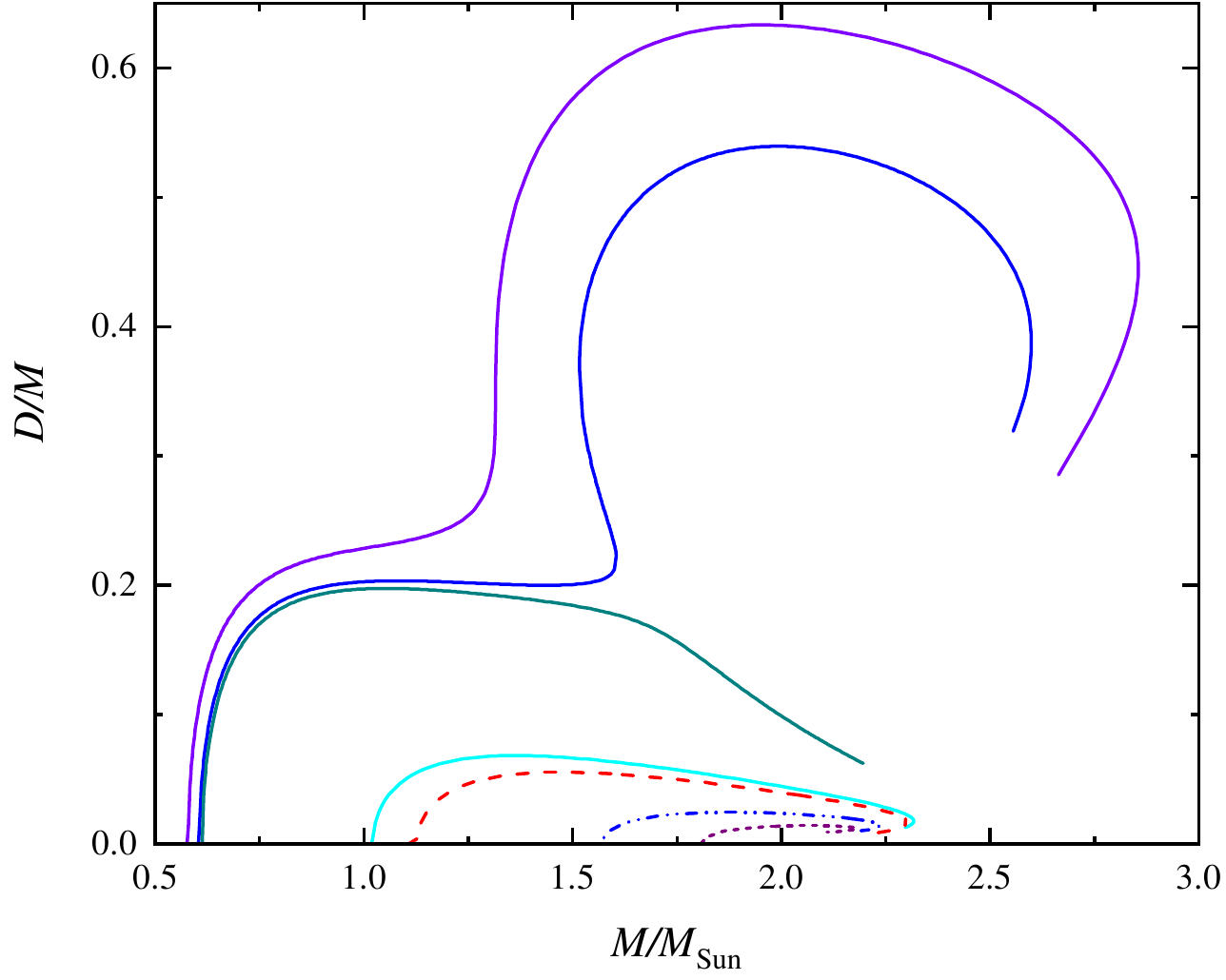}
	\includegraphics[width=0.33\textwidth]{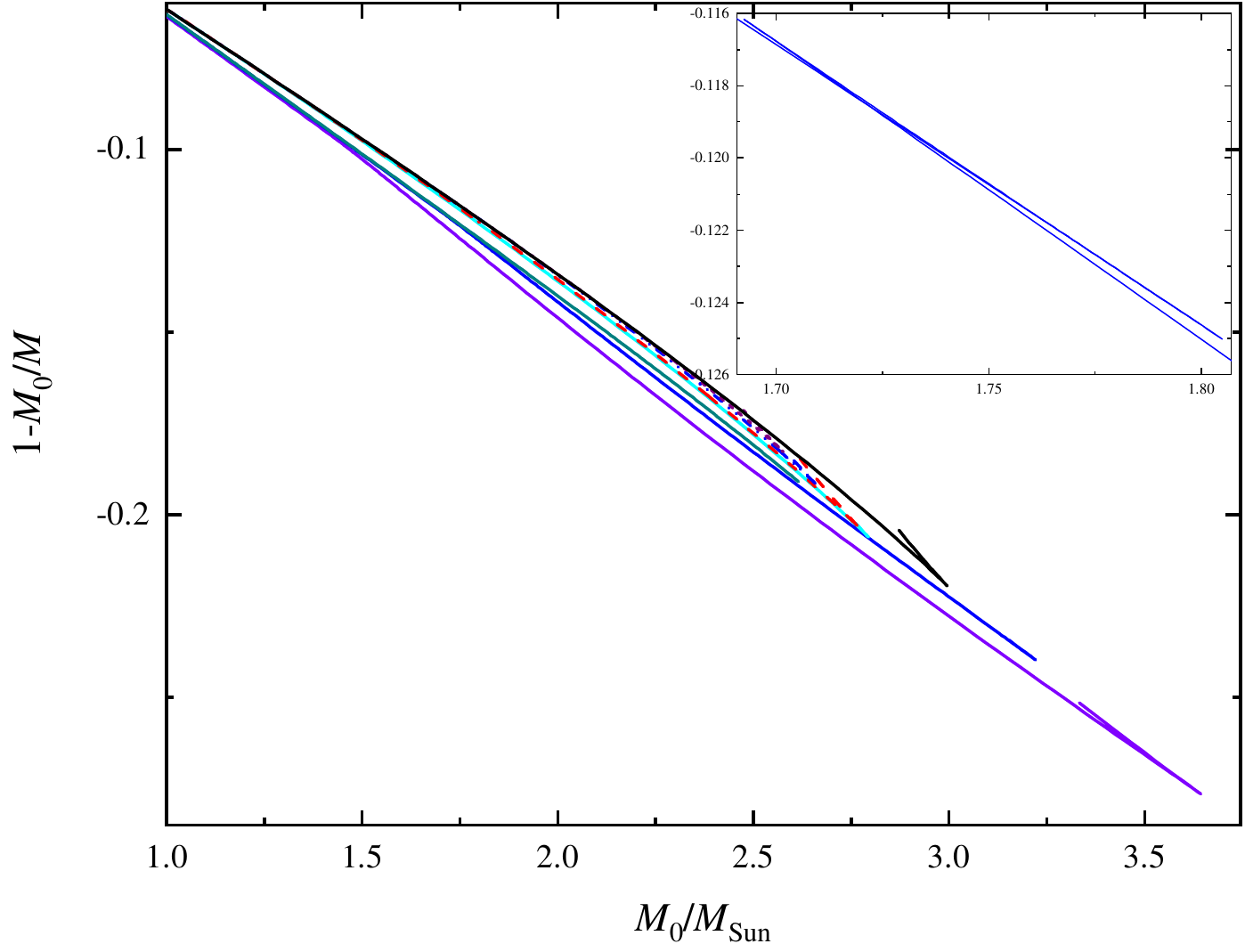}
	\caption{CASE ``sGB-EdGB $+$ STT-DEF'': \textit{Left:} Mass of the star as a function of its radius. \textit{Middle:} The scalar charge of the star, normalized to the mass of the star, as a function of mass. \textit{Right:} The binding energy as a function of the baryon mass. }
	\label{fig:c1_A1}
\end{figure}

We consider a theory with the following combination of coupling functions:
\begin{equation}
	f_{\rm SS}(\varphi)=  \pm \frac{1}{2\beta_{\rm SS}} \left[1-e^{-\beta_{\rm SS}\varphi^2}\right], \,\,\,\,\,\,\, A_{\rm DEF}(\varphi)=e^{\frac{1}{2} \gamma_{\rm DEF} \varphi^2}
\end{equation}

In this case, GR with $\varphi=0$ is always a solution of the field equations because both the sGB and the STT limits of the theory exhibit spontaneous scalarization. The results for the ``$-$'' sign of the coupling function $f_{SS}(\varphi)$ are presented in Fig. \ref{fig:c1_A1}. We plot the mass of the star as a function of its radius in the left panel, the scalar charge normalized to the mass of the star as a function of the mass in the middle panel, and the binding energy in the right panel. 

The parameter $\gamma_{\rm DEF}$ shifts the bifurcation point, and the bigger the absolute value of $\gamma_{\rm DEF}$ is, the further away the bifurcation point is from the sGB one. Larger $\gamma_{\rm DEF}>0$ move the bifurcation point to larger neutron star masses compared to the pure sGB ones, and for $\gamma_{\rm DEF} < 0$ the bifurcation is moved to smaller masses. For larger $\gamma_{\rm DEF}$, when scalarization is only possible due to the sGB term, the maximum mass increase with the increase of $\gamma_{\rm DEF}$ but the branches get shorter due to violation of the regularity condition. For small enough $\gamma_{\rm DEF}$ the STT term also contributes to scalarization and leads to the presence of a ``secondary scalarization'': Below some critical value for $\gamma_{\rm DEF}$ the solutions branch at first is very similar to a sGB branch and at some central energy density it changes to behavior very similar to a DEF branch. There is a short interval of values for $\gamma_{\rm DEF}$ in which this transition happens in a nontrivial way. In that case, the sGB-like branch starts from the bifurcation point and reaches some maximal mass. At this point an intermediate (potentially unstable) branch starts and the mass decreases down to some minimal mass. Once a minimum is reached, a third one, (potentially stable) DEF-like branch originates. The intermediate branch and the presence of a minimum of the mass disappear  when $\gamma_{\rm DEF}$ gets more negative. 

The scalar charge is higher than the pure sGB one for negative $\gamma_{\rm DEF}$ and it is lower than the pure sGB one for positive $\gamma_{\rm DEF}$. In the latter case, significantly lower values of $D/M$ could be reached. The binding energy is always higher, in absolute value, compared to the GR one making the scalarized branches energetically favorable. The binding energy demonstrates very interesting behavior in the cases with negative $\gamma_{\rm DEF}$ and an intermediate branch. In addition to the cusp associated with the maximal mass instability, there are two more cusps. One at the maximal and one at the minimal mass of the intermediate branch. The binding energy of this branch is smaller in absolute value compared to the two main branches. Therefore, it is most probably unstable. This could have important implications: a neutron star that moves along the $M(R)$ curve, will need to make a jump from the sGB-like part of the branch to the DEF part or vice versa. Such a ``move'' can happen during the fallback accretion in stellar collapse, as a supermassive star compactifies and slows down after binary merger, or even in the inspiral pre-merger phase when an effect similar to dynamical scalarization can be observed \cite{Shibata:2013pra,Kuan:2023trn}.

Concerning the $\epsilon = +1$ case, the behavior is very similar to the one described above, though, not identical. In that case, the branches which bifurcate below the pure sGB one stay below the sGB curve, and the branches which bifurcate above the GB one are always above the GB one. No resemblance with the DEF STT can be observed for moderately high values for $\gamma_{\rm DEF}$.

\subsection{``sGB-EdGB $+$ STT-DEF'':  EdGB gravity combined with DEF spontaneous scalarization model.}

\begin{figure}[htb]
	\includegraphics[width=0.33\textwidth]{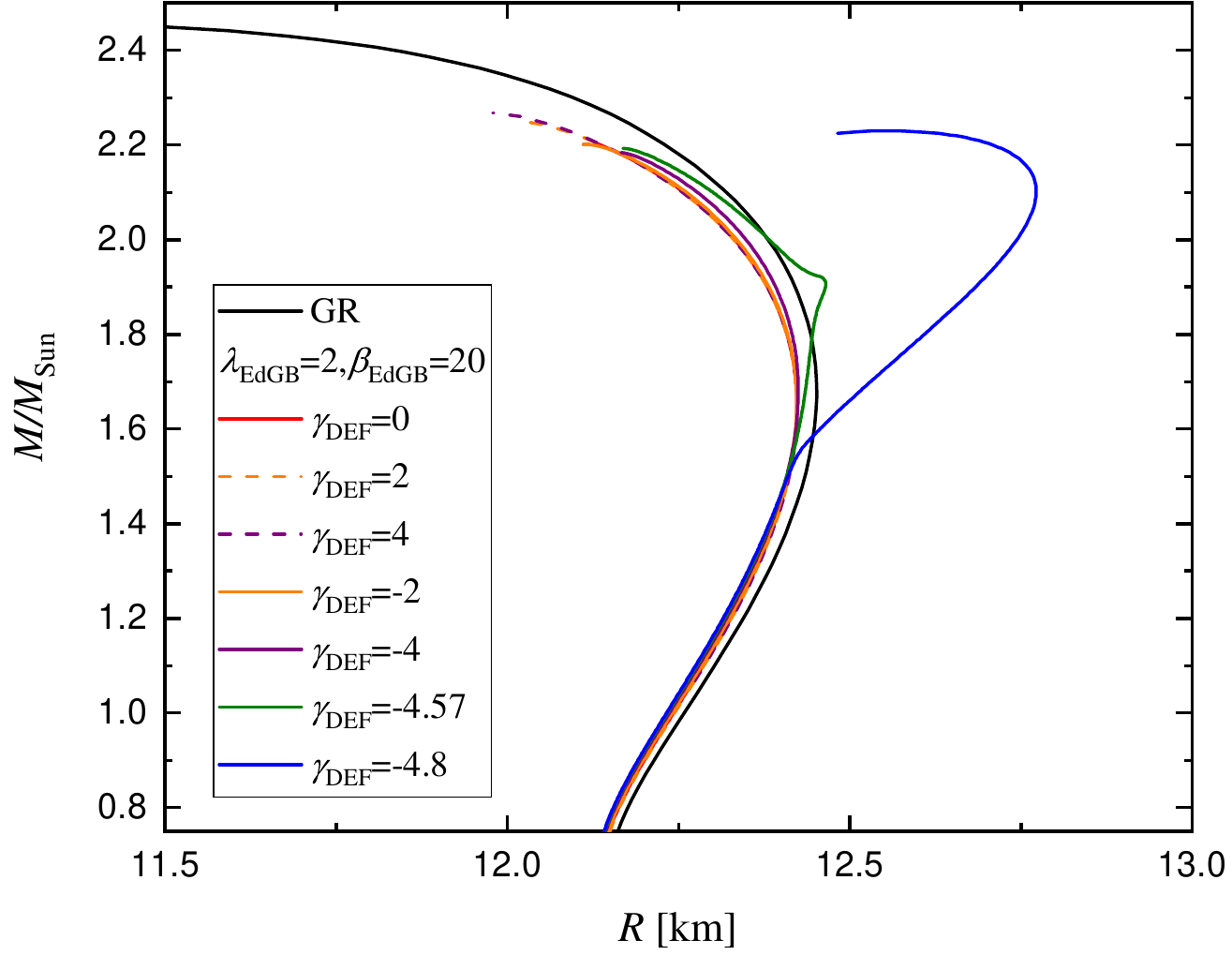}
	\includegraphics[width=0.33\textwidth]{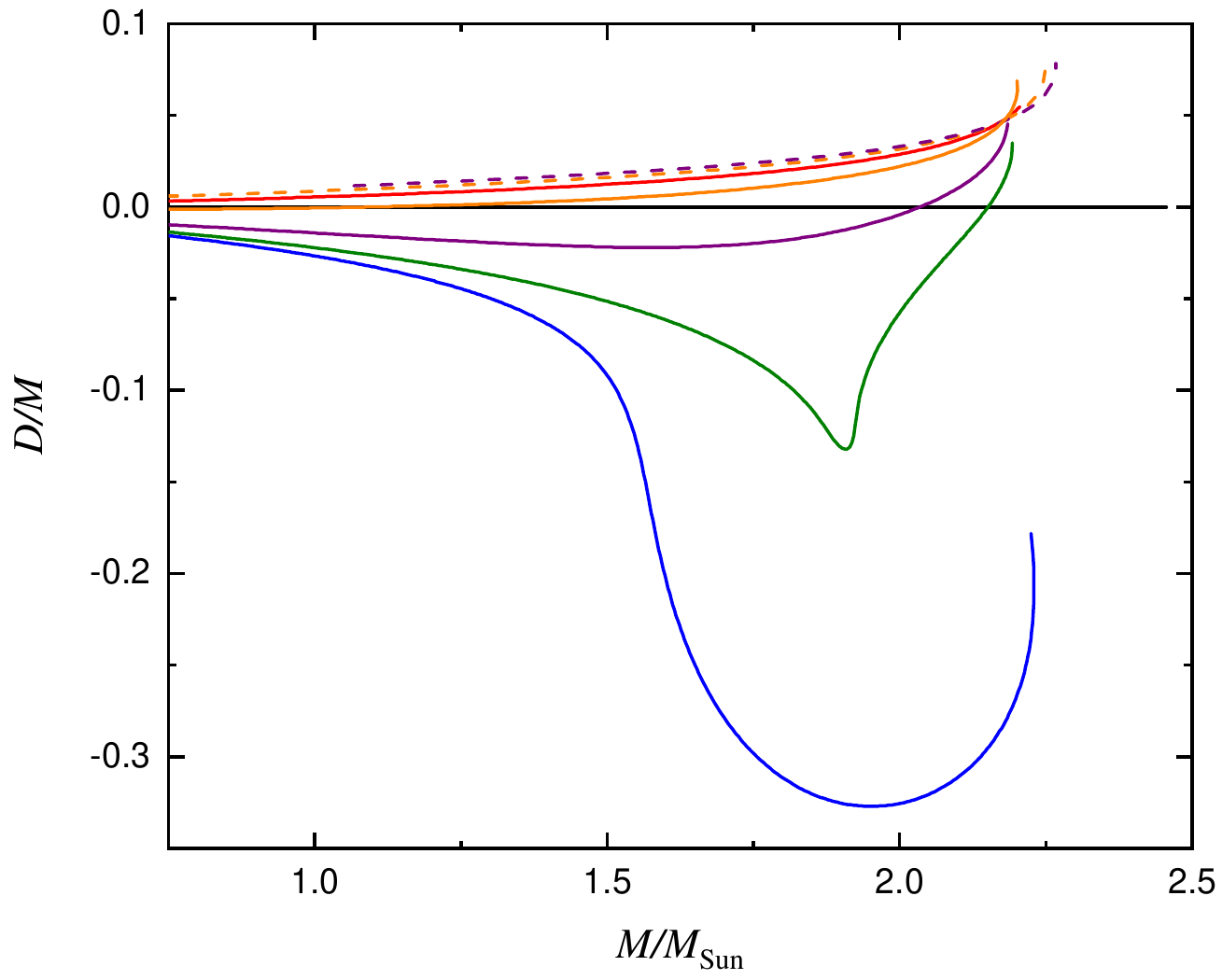}
	\includegraphics[width=0.33\textwidth]{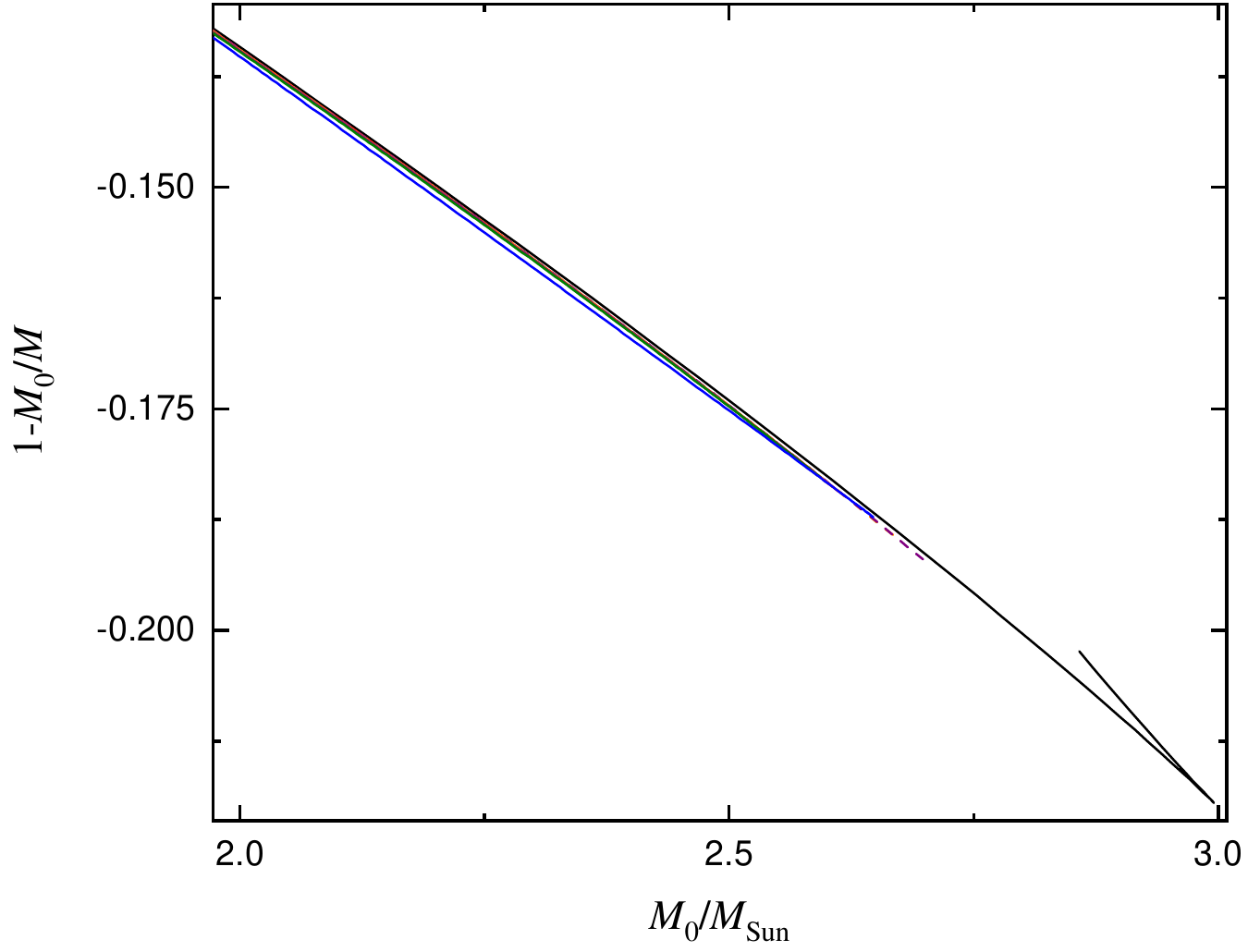}
	\caption{CASE ``sGB-EdGB $+$ STT-DEF'':  \textit{Left:} Mass of the star as a function of its radius. \textit{Middle:} The scalar charge of the star, normalized to the mass of the star, as a function of mass. \textit{Right:} The binding energy as a function of the baryon mass.}
	\label{fig:expGB_A1}
\end{figure}

We consider a theory with the following combinations of coupling functions:
\begin{equation}
	f_{\rm EdGB}(\varphi) = \frac{1}{2\beta_{\rm EdGB}}e^{2\beta_{\rm EdGB} \varphi}, \,\,\,\,\,\, A_{\rm DEF}(\varphi)=e^{\frac{1}{2} \gamma_{\rm DEF} \varphi^2} .
\end{equation}

The results for the coupling functions under consideration are presented in Fig. \ref{fig:expGB_A1}.  Both positive and negative values for $\gamma_{\rm DEF}$ are studied. For positive values, where the pure DEF models exhibit no scalarization and the neutron stars coincide with GR (see \cite{Mendes:2016fby} for an exception), the branches get longer compared to the pure sGB case, but no other deviations are observed. For $\gamma_{\rm DEF} < 0$, though, behavior very similar to the spontaneous scalarization in the pure DEF STT can be observed. With the decrease of $\gamma_{\rm DEF}$, at first there are very small deviations from the pure sGB branch. Below some critical value for $\gamma_{\rm DEF}$, for which a scalarization in the pure DEF case exist, the branches exhibit interesting behavior. At first, the solution branch follows closely the pure EdGB branch until a given central energy density at which the branch makes a sharp turn to higher radii. This is similar to the secondary scalarization discussed in the previous section.  The high mass, DEF-like part of the branch reaches higher masses as $\gamma_{\rm DEF}$ gets more negative. 

An interesting behavior of the scalar charge is observed for small enough $\gamma_{\rm DEF} <0$, namely $D/M$ changes its sign. Therefore, there is a part of the branch for which the scalar charge is vanishing small that could have implications when studying the observational constraints.  Concerning the binding energy, it is always larger by absolute value compared to the GR case and the maximum mass marks a change of stability.

\section{Conclusions}

In the present paper, we constructed static spherically symmetric neutron stars in extended scalar-tensor theories (ESTT) which can be considered as a nonlinear combination of the standard sGB gravity and the classical scalar-tensor theories. Such combinations are interesting and explored in the context of nonlinear simulations \cite{Evstafyeva:2022rve} since they can lead to the appearance of interesting effects such as breathing modes.  We make an exhaustive study of various combinations of the Gauss-Bonnet and scalar-tensor theory  couplings. As it turns out, in certain cases, interesting qualitative changes in the solution spectrum can arise. One of the main reasons for that is the qualitatively very different behavior of the two limiting cases -- for pure sGB gravity the neutron stars are more compact than in GR and they reach lower maximum masses, while it is exactly the other way around in the classical scalar-tensor theories.

Depending on the specific combination of Gauss-Bonnet and scalar-tensor theory couplings, we observe different families of solutions. In some cases, qualitatively new branches of neutron stars appear. For example, in an ESTT being a combination between sGB with scalarization and Brans-Dicke theory, some of the new solution branches are terminated at a minimal mass below which no compact stars exist. Based on binding energy studies, we have determined that most probably two of these branches are stable and coexist in a certain density interval. Which one would realize in practice is a difficult question that can be answered by performing nonlinear time evolution. 

In another case, namely for an ESTT where the scalar-tensor theory sector exhibits spontaneous scalarization (DEF model) combined with the Einstein-dilaton-Gauss-Bonnet gravity, a nontrivial shape of the solution branches is observed such as the appearance of sharp bumps. This normally happens in the transition region where the effect of sGB starts dominating over the DEF contribution or the other way around. This effect can not be mimicked by a simple equation of state change in GR and it shares some similarities with the effects produced by equations of state exhibiting phase transitions.

An important quantity we studied in all cases is the ratio between the scalar charge of the models and their masses, $D/M$. This ratio is directly related to the emission of scalar dipole radiation from a binary system and can be constrained by observations such as binary pulsars and the pre-merger inspiral. The behavior of the scalar charge along a sequence of solutions can be highly nontrivial, significantly different from the pure sGB and scalar-tensor theory cases. Another interesting fact is that for some solutions branches the scalar charge can flip sign. This means that even if the majority of the models along a given branch have high scalar charges, there will be some models with very low and even zero scalar charge. This could have an impact on the theory constraint and we plan to investigate it in future work.

\section*{Acknowledgements}
This study is in part financed by the European Union-NextGenerationEU, through the National Recovery and Resilience Plan of the Republic of Bulgaria, project No. BG-RRP-2.004-0008-C01. DD  acknowledges financial support via an Emmy Noether Research Group funded by the German Research Foundation (DFG) under grant no. DO 1771/1-1.

\appendix

\section{``sGB-shift-sym $+$ STT-DEF'':  Shift symmetric sGB gravity combined with DEF spontaneous scalarization model.} \label{app:Sym_DEF}
\begin{figure}[htb]
	\includegraphics[width=0.33\textwidth]{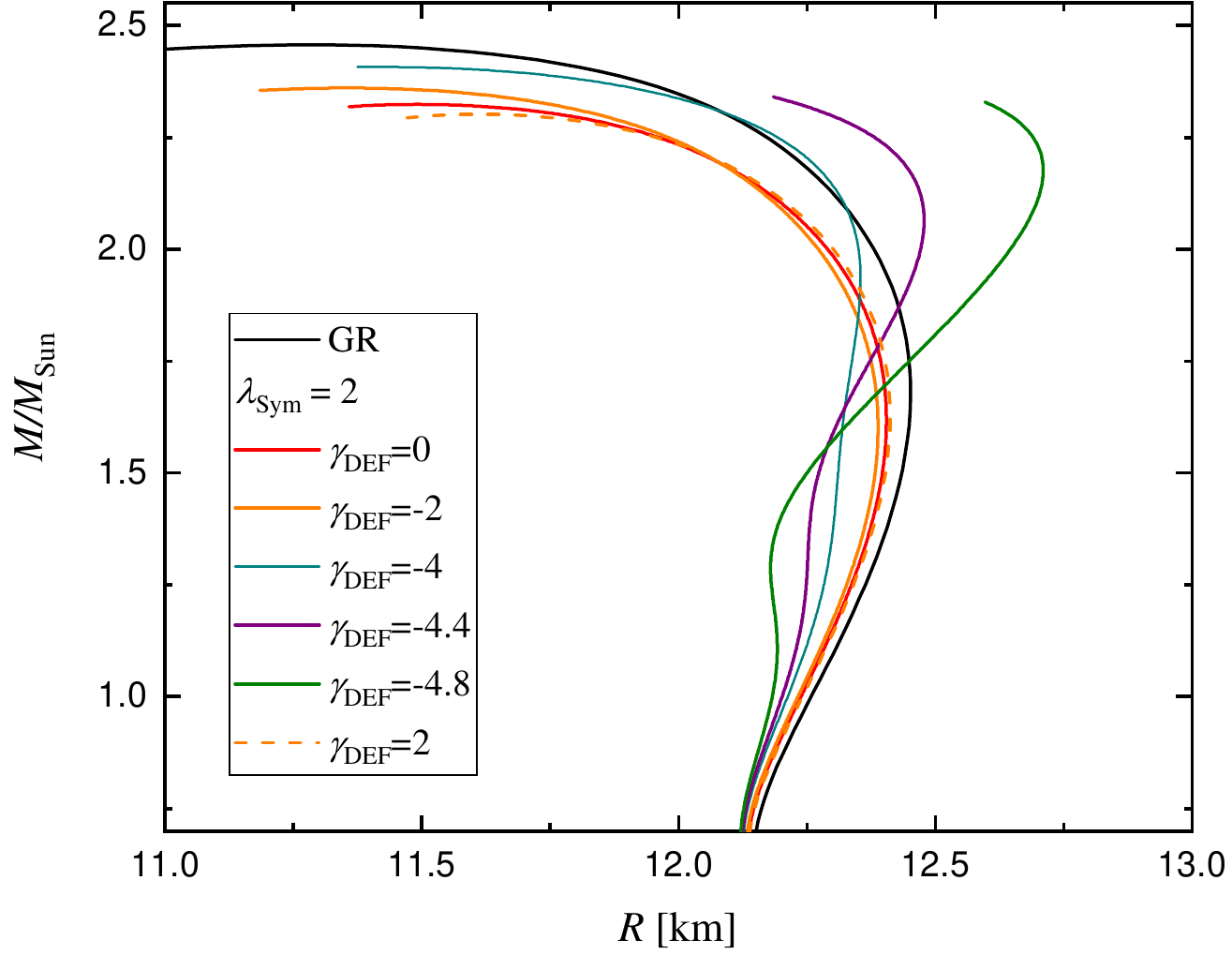}
	\includegraphics[width=0.33\textwidth]{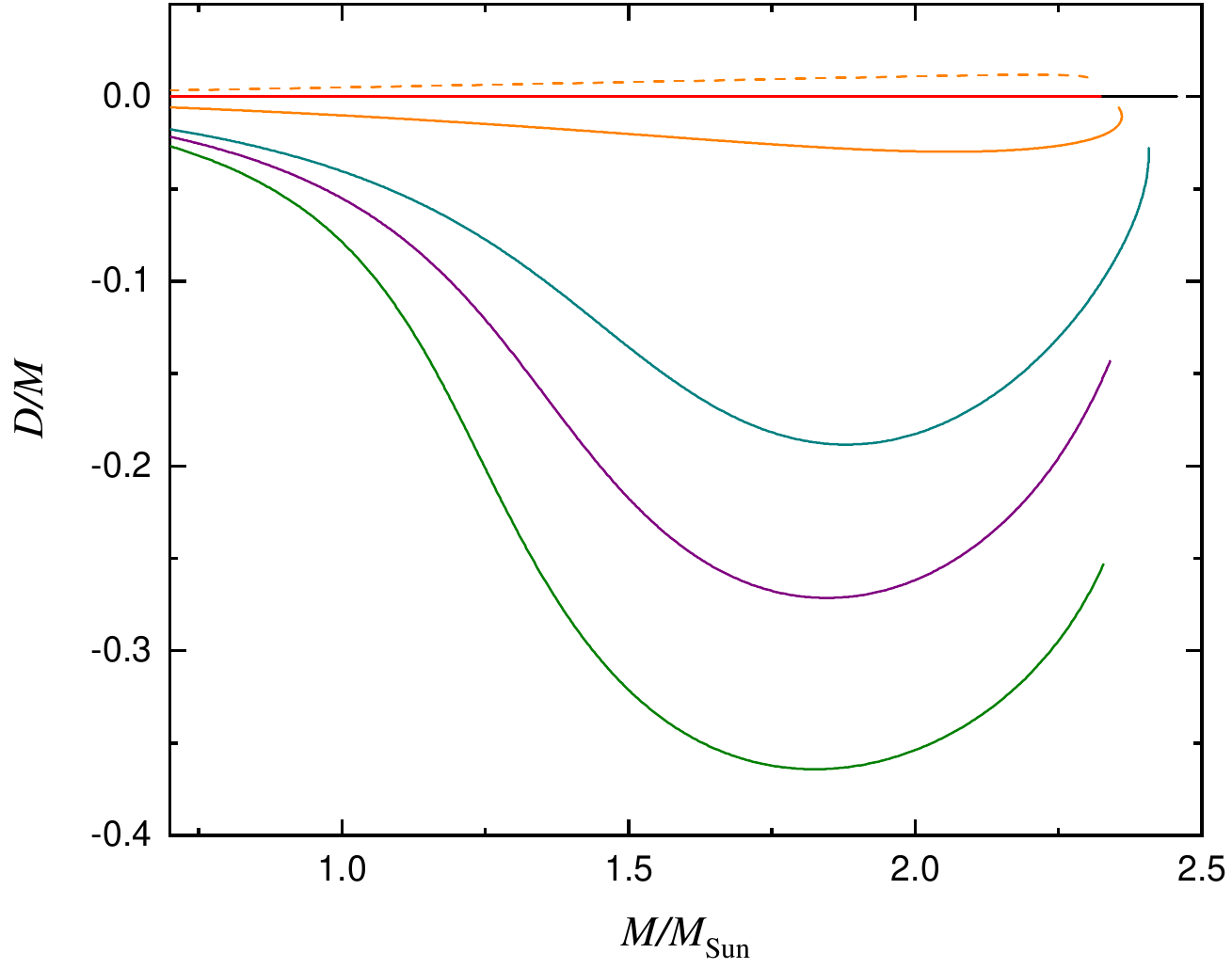}
	\includegraphics[width=0.33\textwidth]{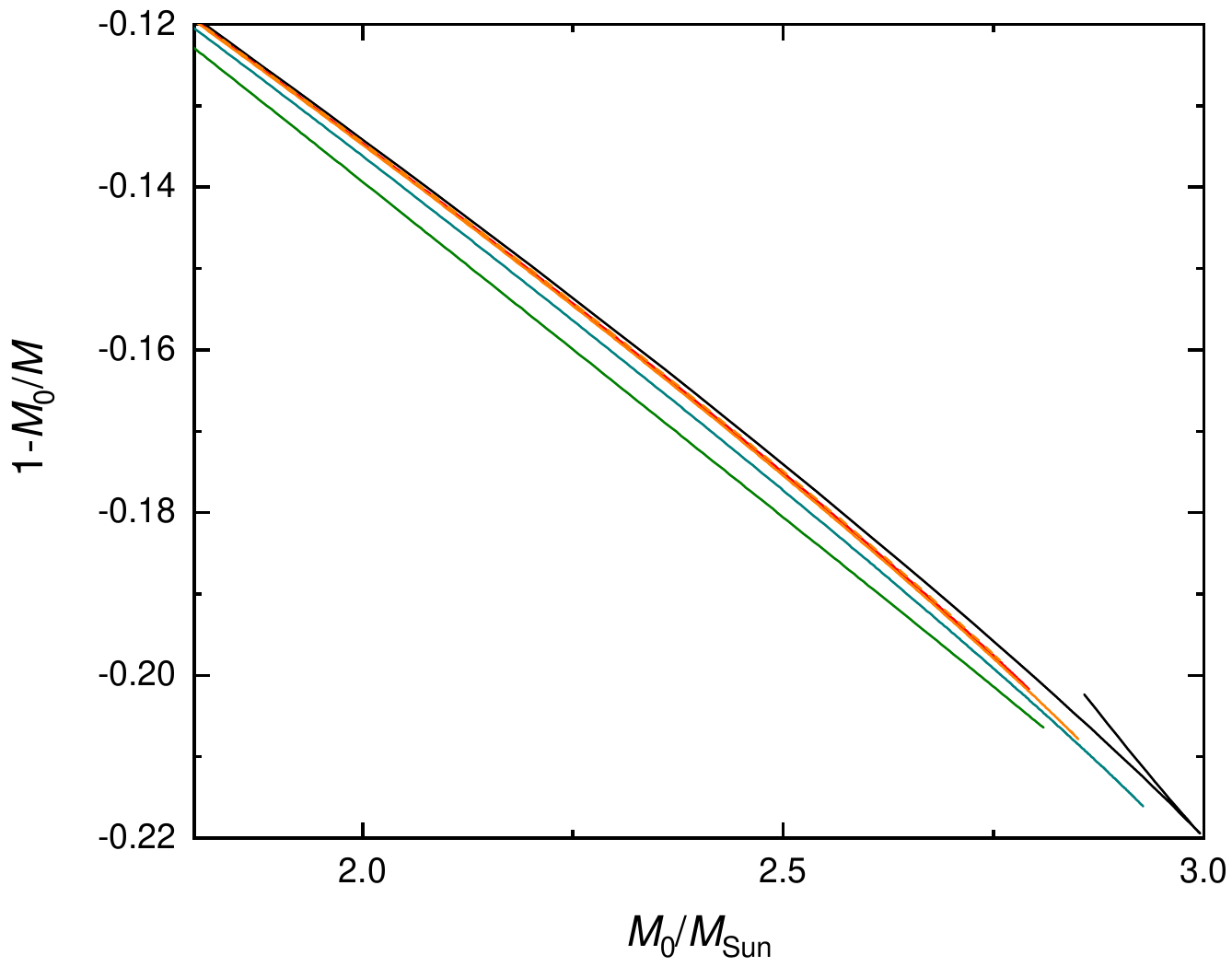}
	\caption{CASE ``sGB-shift-sym $+$ STT-DEF'': \textit{Left:} Mass of the star as a function of its radius. \textit{Middle:} The scalar charge of the star, normalized to the mass of the star, as a function of mass. \textit{Right:} The binding energy as function of the baryon mass.}
	\label{fig:shift_GB_A1}
\end{figure}

We consider a theory with the following combinations of coupling functions:
\begin{equation}
	f_{\rm Sym}(\varphi)=\varphi, \,\,\,\,\,\, A_{\rm DEF}(\varphi)=e^{\frac{1}{2} \gamma_{\rm DEF} \varphi^2}
\end{equation}
The shift-symmetric sGB part can be viewed as a weak coupling limit of the EdGB gravity with exponential coupling, therefore similarities with the ``sGB-EdGB $+$ STT-DEF'' case are expected. 
For $\gamma_{\rm DEF}= 0$ we recover the shift-symmetric sGB gravity in which, as well known \cite{Yagi:2015oca}, the neutron stars have no scalar charge, therefore, no constraints can be put on the theory by the observations of the orbital decay of binary pulsars. The presence of a coupling $A_{\rm DEF}(\varphi)$ with $\gamma_{\rm DEF}\ne 0$, though, as we will demonstrate in this section, introduces nonvanishing scalar charge.  

In the left panel of Fig. \ref{fig:shift_GB_A1} we present mass of radius relations for neutron stars in GR, in pure sGB gravity (i.e. $\gamma_{\rm DEF} = 0$), and in 
the extended scalar tensor theory under consideration  for different values for the parameter $\gamma_{\rm DEF}$, both positive and negative.  For $\gamma_{\rm DEF} > 0$ the deviations from pure sGB are both qualitatively and quantitatively very small. For $\gamma_{\rm DEF}  < 0$ the picture is rather different (even for values of $\gamma_{\rm DEF}  < 0$ where scalarization in the pure DEF case is not possible). For intermediate masses, the radius of the neutron star decreases, compared to the pure sGB one. For higher masses, and small enough $\gamma_{\rm DEF}$, the radius increases beyond the pure sGB one and even beyond the GR one, resembling DEF-like scalarization. We can think of this as some form of induced secondary scalarization. The maximum neutron star mass increase above the sGB one and even above the GR one for negative enough values of $\gamma_{\rm DEF}$.

In the middle panel of Fig. \ref{fig:shift_GB_A1} we present the scalar charge normalized to the mass of the star.  The pure (shift-symmetric) sGB has no scalar charge. The presence of the matter conformal  coupling introduces non-vanishing scalar charge. It is still relatively small for $\gamma_{\rm DEF}<0$  and increases significantly for negative $\gamma_{\rm DEF}$, resembling the one in pure DEF model. The binging energy exhibits a cusp, a change of stability, only at the maximum neutron star mass of each sequence.

It is interesting to point out that there is a qualitative difference with the ``sGB-EdGB $+$ STT-DEF'' case presented in Fig. \ref{fig:c1_A1}, namely an exponential GB coupling instead of a shift-symmetric one. In the DEF case, and for $\gamma_{\rm DEF} < 0$,  the transition from sGB-led to DEF-led branch happens sharply. In the shift-symmetric case we have fluent transitions as $\gamma_{\rm DEF}$ gets more negative and a significant deviation from pure sGB even for small neutron star masses.

\section{``sGB-shift-sym $+$ STT-BD'': Shift-symmetric sGB gravity combined with Brans-Dicke scalar-tensor theory} \label{app:Sym_BD}

\begin{figure}[htb]
	\includegraphics[width=0.33\textwidth]{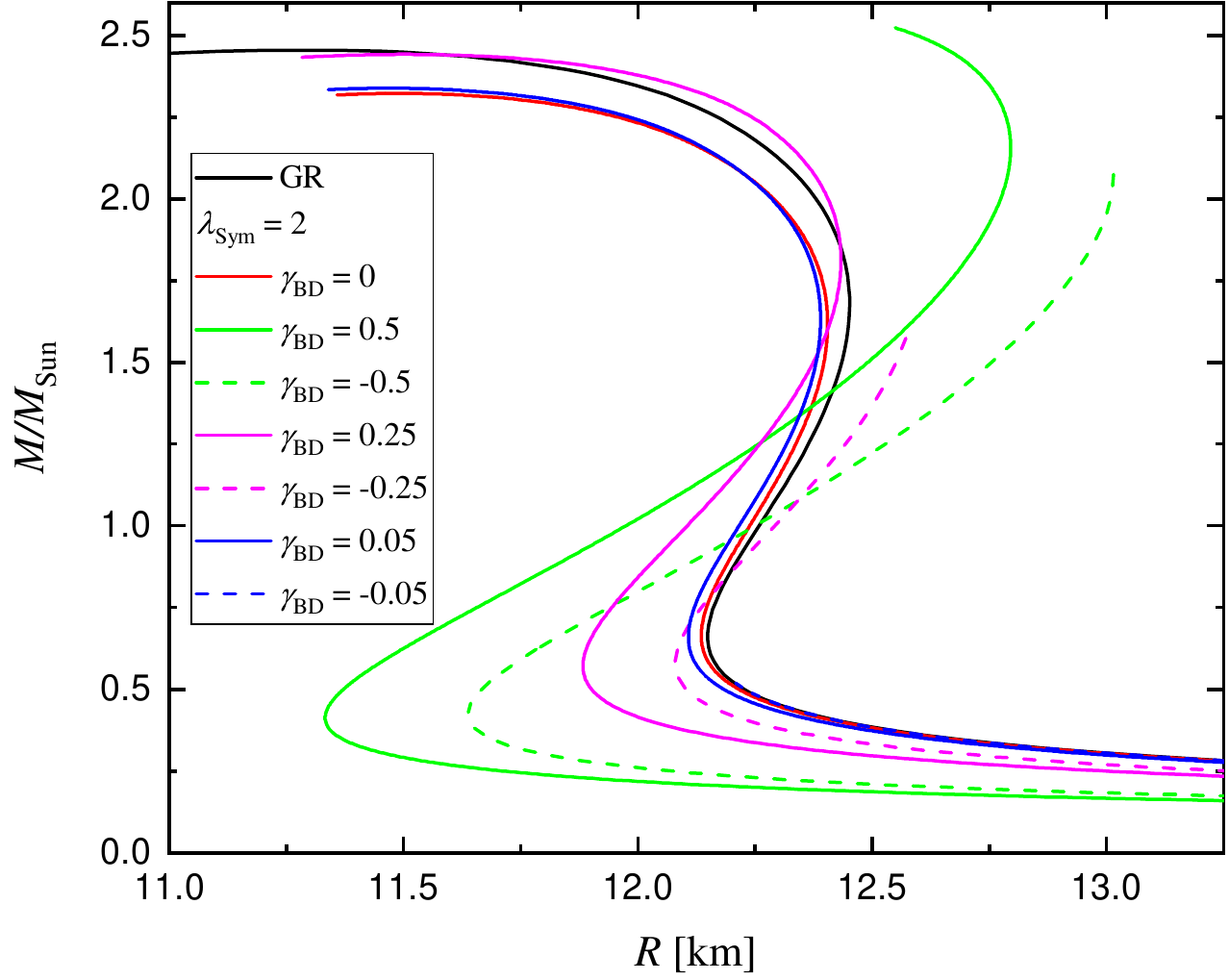}
	\includegraphics[width=0.33\textwidth]{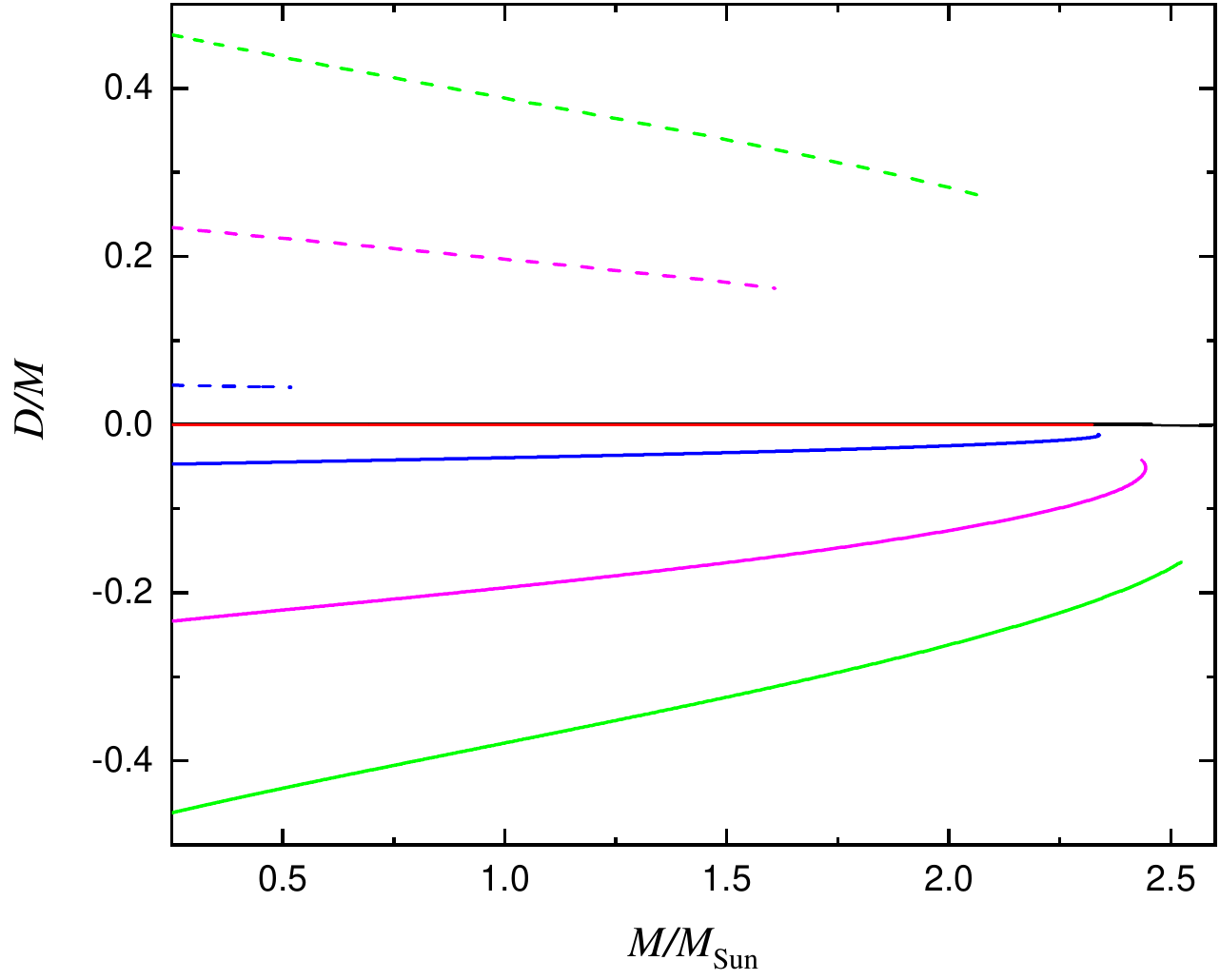}
	\includegraphics[width=0.33\textwidth]{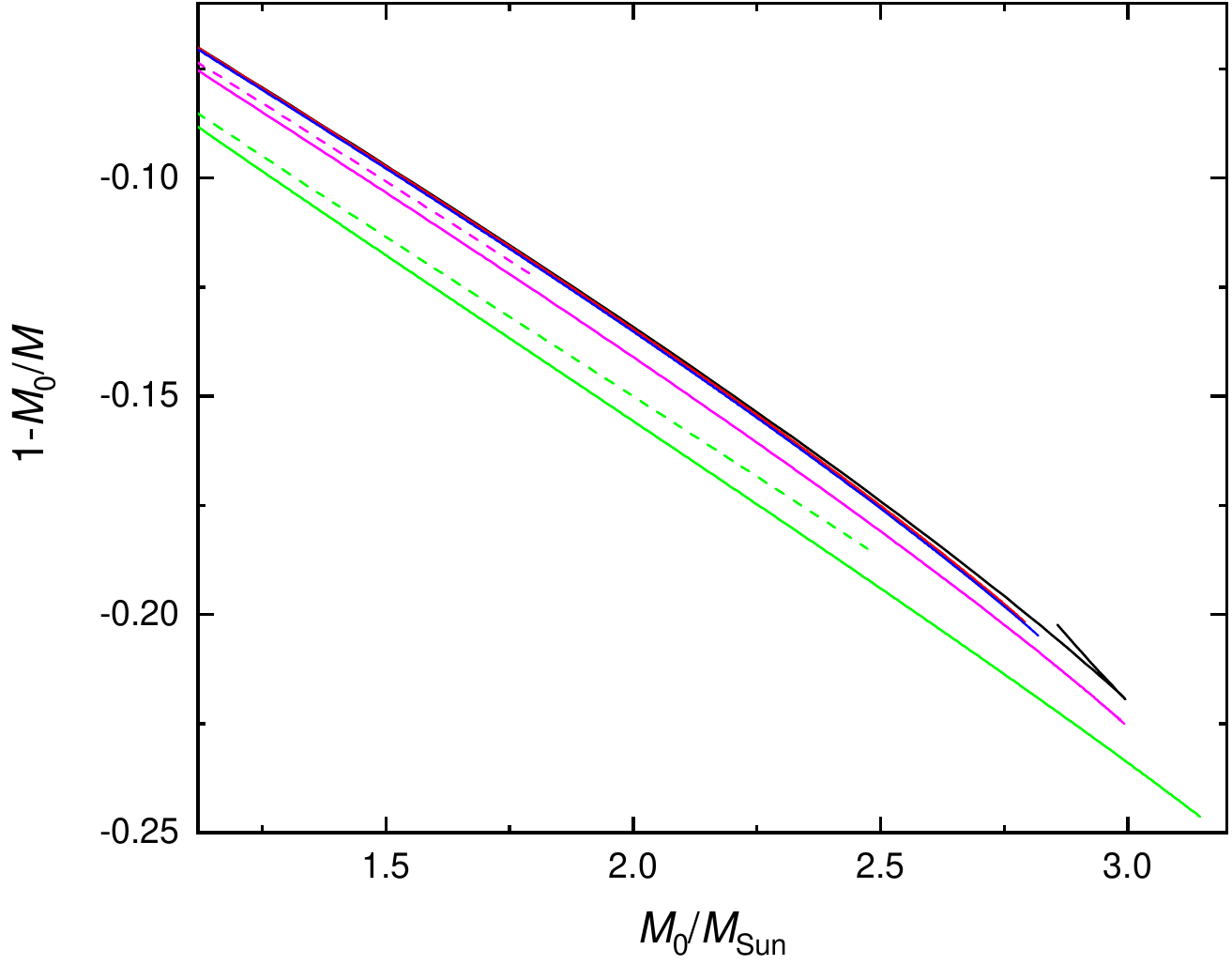}
	\caption{ CASE ``sGB-shift-sym $+$ STT-BD'':  \textit{Left:} Mass of the star as a function of its radius. \textit{Middle:} The scalar charge of the star, normalized to the mass of the star, as a function of mass. \textit{Right:} The binding energy as a function of the baryon mass.}
	\label{fig:shift_GB_A2}
\end{figure}

We consider a theory with the following combinations of coupling functions:
\begin{equation}
	f_{\rm Sym}(\varphi)=\varphi, \,\,\,\,\,\, A_{\rm BD}(\varphi)=e^{ \gamma_{\rm BD} \varphi} 
\end{equation}
The results are presented in Fig. \ref{fig:shift_GB_A2}.  In the $M(R)$ relation for $\gamma_{\rm BD}>0$ the maximal masses increase above the pure sGB and even above GR with the increase of $\gamma_{\rm BD}$. As $\gamma_{\rm BD}$ grows, the radius of the models with higher masses increases significantly, while, the radius of the models with low masses decreases. This behavior is reminiscent of the pure BD case. Concerning the normalized scalar charge $\frac{D}{M}$, it increases with $\gamma_{\rm BD}$, and the binding energy increases by absolute value compared to the GR one.

In the $\gamma_{\rm BD} < 0$ case the branches of solutions have the same shape as their positive counterparts. In this case, though, the radius of all models is higher and the branches are terminated for lower masses. The binding energy is lower, compared to the $\gamma_{\rm BD}>0$ case.

\section{``sGB-EdGB $+$ STT-BD'': EdGB gravity combined with Brans-Dicke scalar-tensor theory} \label{app:EdGB_BD}

\begin{figure}[htb]
	\includegraphics[width=0.33\textwidth]{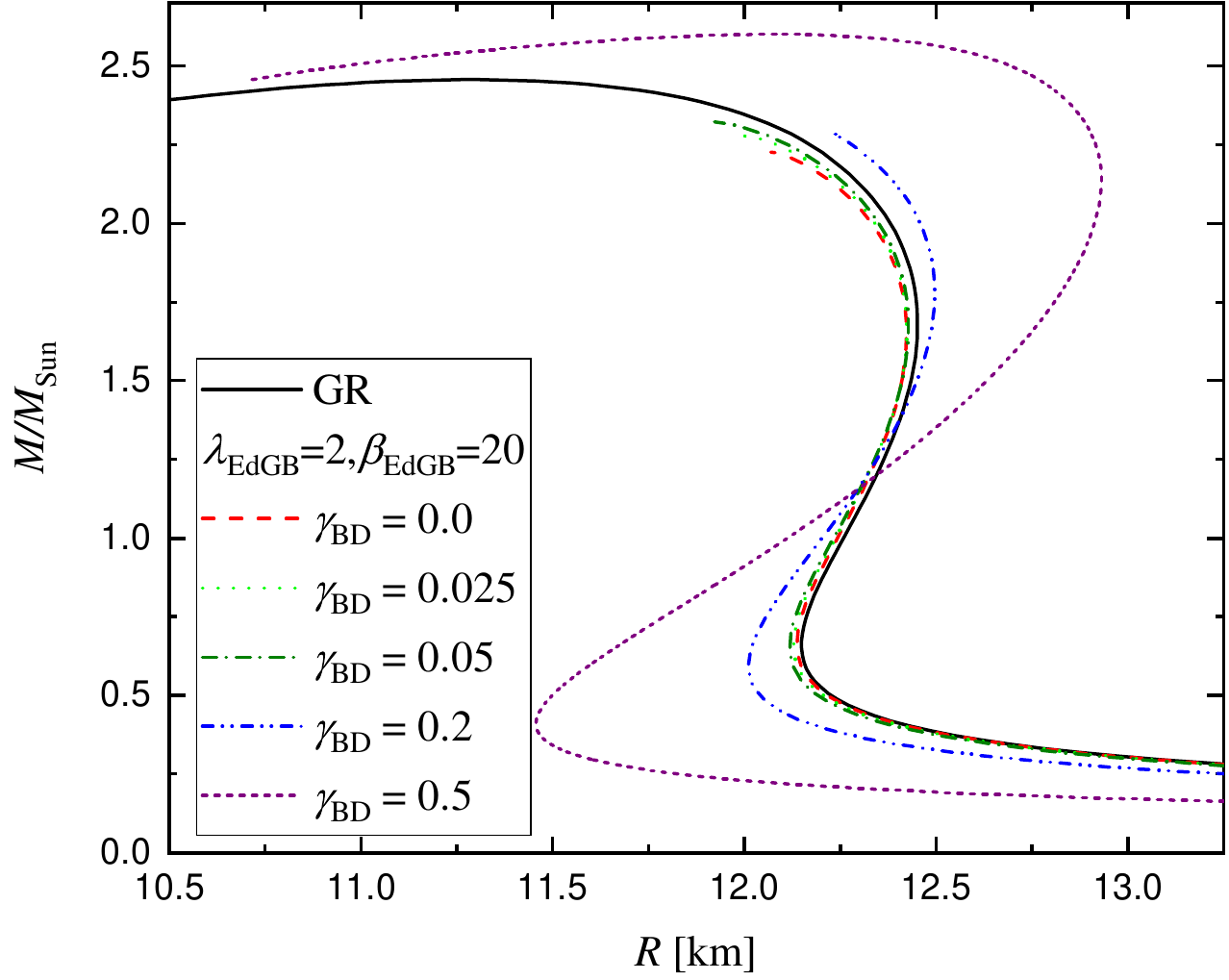}
	\includegraphics[width=0.33\textwidth]{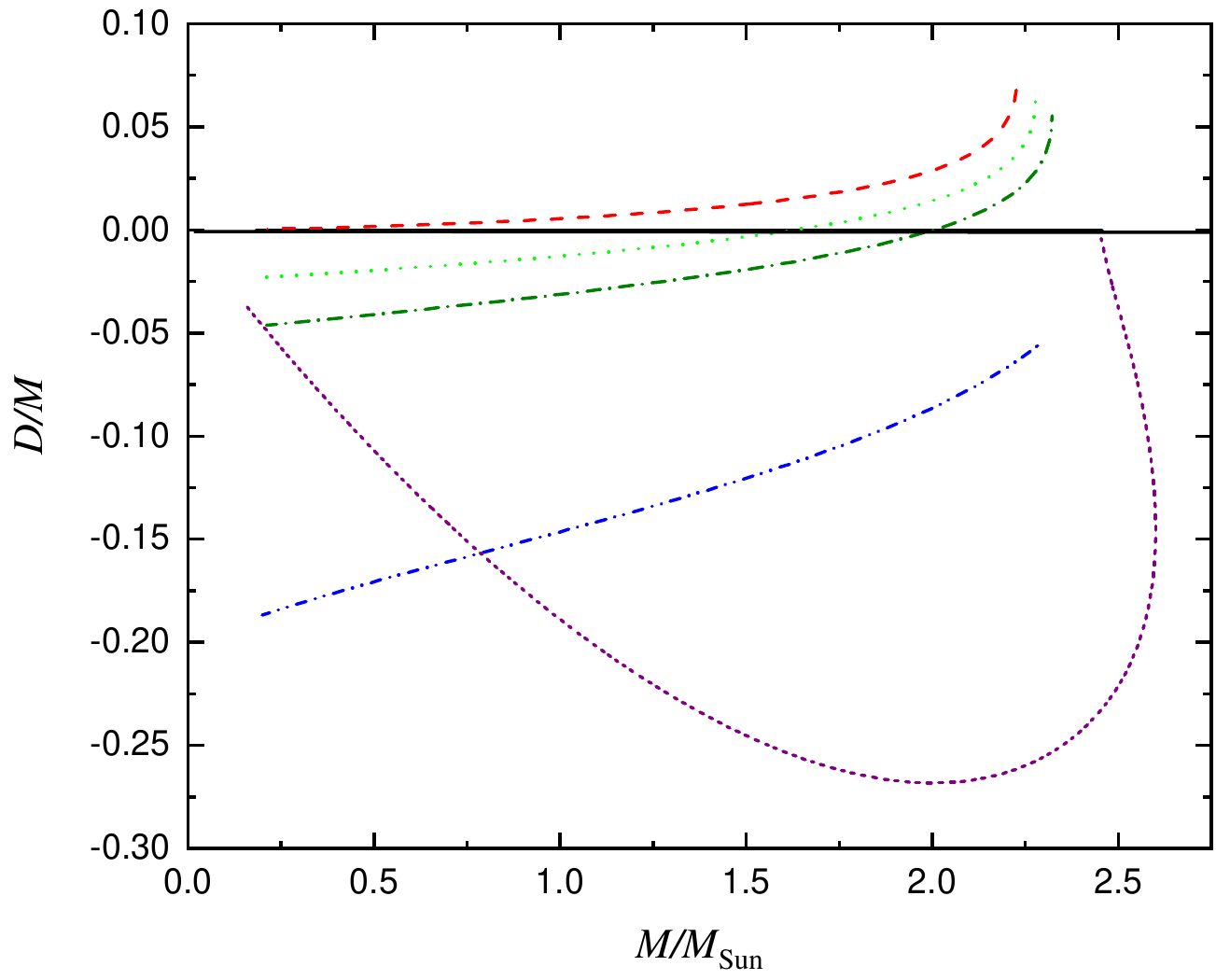}
	\includegraphics[width=0.33\textwidth]{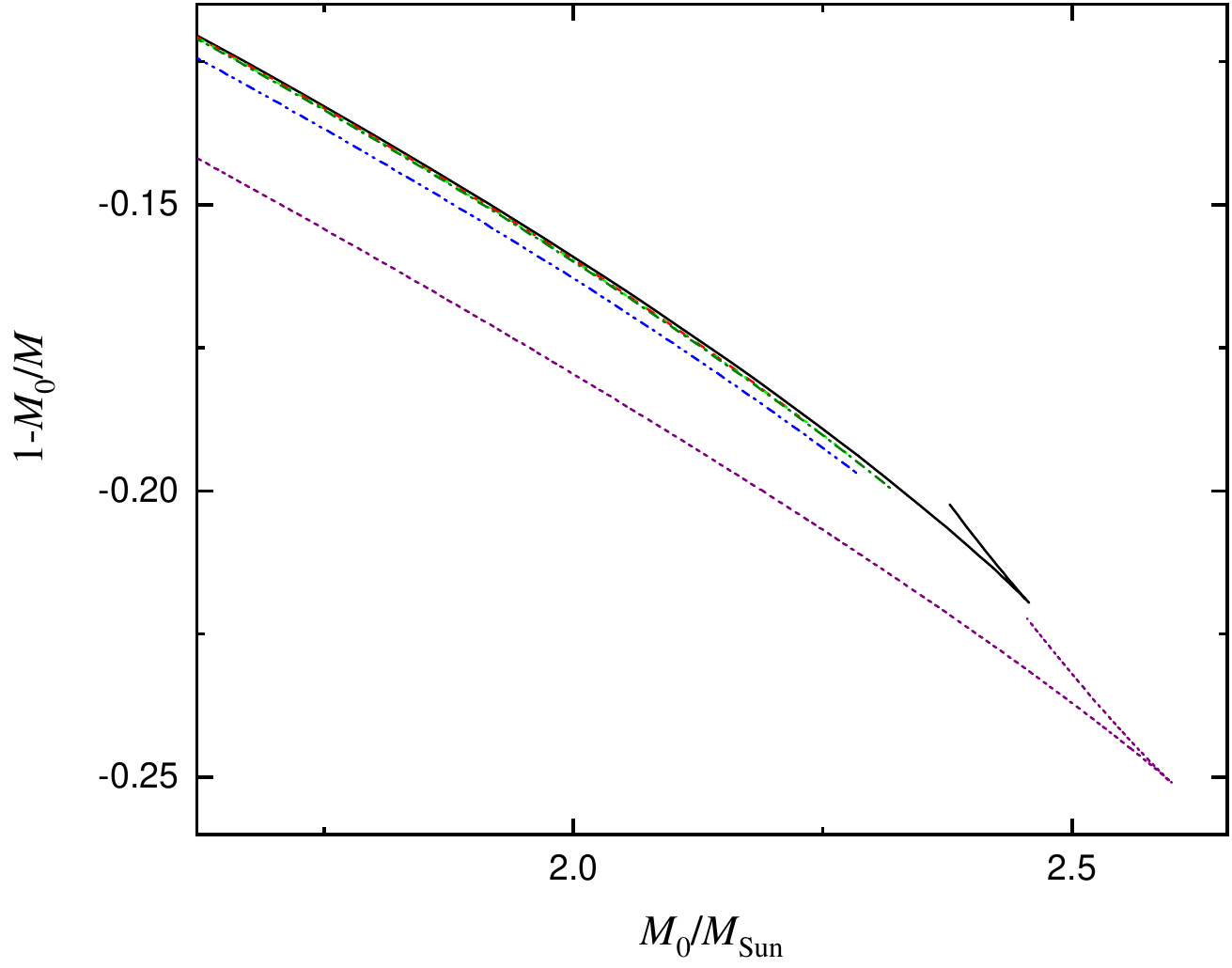}
	\caption{CASE ``sGB-EdGB $+$ STT-BD'': \textit{Left:} Mass of the star as a function of its radius. \textit{Middle:} The scalar charge of the star, normalized to the mass of the star, as a function of mass. \textit{Right:} The binding energy as a function of the baryon mass. }
	\label{fig:expGB_A2}
\end{figure}

We consider a theory with the following combinations of coupling functions:
\begin{equation}
	f_{\rm EdGB}(\varphi) = \frac{1}{2\beta_{\rm EdGB}}e^{2\beta_{\rm EdGB} \varphi}, \,\,\,\, A_{\rm BD}(\varphi)=e^{  \gamma_{\rm BD} \varphi} 
\end{equation}

The numerical results for this case are presented in  Fig. \ref{fig:expGB_A2}. The observed behavior is qualitatively very similar to the CASE ``sGB-shift-sym $+$ STT-BD''. That is why we will only present the figures for $\gamma_{\rm BD} > 0$ without commenting on them. The only qualitative difference with respect to CASE ``sGB-shift-sym $+$ STT-BD'' is that for some values of $\gamma_{\rm BD}$ the scalar charge changes its sign.


\bibliography{references}

\end{document}